\documentclass[aps,prb,superscriptaddress,twocolumn,floatfix]{revtex4-1}

\usepackage{graphicx,graphics}
\usepackage{dcolumn}
\usepackage{amsmath,amssymb,amsfonts}
\usepackage{latexsym,verbatim}
\usepackage{bm}
\usepackage{color}
\usepackage{ulem}
\usepackage[percent]{overpic}
\usepackage[breaklinks=true,colorlinks,citecolor=blue,linkcolor=blue,urlcolor=blue]{hyperref}
\begin{document}
\title{Generation and morphing of plasmons in graphene superlattices}
\author{Andrea Tomadin}
\email{andrea.tomadin@sns.it}
\affiliation{NEST, Istituto Nanoscienze-CNR and Scuola Normale Superiore, I-56126 Pisa,
Italy}
\author{Francisco Guinea}
\affiliation{Instituto de Ciencia de Materiales de Madrid (CSIC), Sor Juana In\'es de la Cruz 3, E-28049 Madrid, Spain}
\author{Marco Polini}
\affiliation{NEST, Istituto Nanoscienze-CNR and Scuola Normale Superiore, I-56126 Pisa,
Italy}
\affiliation{Istituto Italiano di Tecnologia, Graphene Labs, Via Morego 30, I-16163 Genova, Italy}
\begin{abstract}
Recent experimental studies on graphene on hexagonal Boron Nitride (hBN) have demonstrated that hBN is not only a passive substrate that ensures superb electronic properties of graphene's carriers, but that it actively modifies their massless Dirac fermion character through a periodic moir\'e potential. In this work we present a theory of the plasmon excitation spectrum of massless Dirac fermions in a moir\'e superlattice. We demonstrate that graphene-hBN stacks offer a rich platform for plasmonics in which control of plasmon modes can occur not only via electrostatic gating but also by adjusting e.g. the relative crystallographic alignment.
\end{abstract}

\maketitle

\section{Introduction}
\label{sect:intro}

Vertical heterostructures~\cite{novoselov_ps_2012,bonaccorso_matertoday_2012,geim_nature_2013} comprising graphene and two-dimensional (2D) hexagonal Boron Nitride (hBN) crystals~\cite{dean_naturenano_2010} offer novel opportunities for applications~\cite{britnell_science_2012} and fundamental studies of electron-electron interactions~\cite{elias_naturephys_2011,yu_pnas_2013,gorbachev_naturephys_2012}. Recent experimental studies~\cite{yankowitz_naturephys_2012,ponomarenko_nature_2013,dean_nature_2013,hunt_science_2013} have demonstrated that hBN substantially alters the electronic spectrum of the massless Dirac fermion (MDF) carriers hosted in a nearby graphene sheet. Indeed, when graphene is deposited on hBN, it displays a moir\'e pattern~\cite{xue_naturemater_2011,decker_nanolett_2011}, a modified tunneling density of states~\cite{yankowitz_naturephys_2012}, and self-similar transport characteristics in a magnetic field~\cite{ponomarenko_nature_2013,dean_nature_2013,hunt_science_2013}.
The potential produced by hBN acts on graphene's carriers as a perturbation with the periodicity of the moir\'e pattern~\cite{graphenesuperlattices,wallbank_prb_2013}.
This is responsible for a reconstruction of the MDF spectrum and the emergence of minibands in the moir\'e superlattice Brillouin zone (SBZ)~\cite{graphenesuperlattices,wallbank_prb_2013}.

A parallel line of research has focussed a great deal of attention on graphene plasmonics~\cite{grapheneplasmonics}. Here, the goal is to exploit the interaction of infrared light with ``Dirac plasmons" (DPs)---the self-sustained density oscillations of the MDF liquid in a doped graphene sheet~\cite{Diracplasmons}---for a variety of applications such as infrared~\cite{freitag_naturecommun_2013} and Terahertz~\cite{vicarelli_naturemater_2012} photodetectors. Interest in graphene plasmonics considerably increased after two experimental groups~\cite{fei_nature_2012,chen_nature_2012} showed that the DP wavelength is much smaller than the illumination wavelength, allowing an extreme concentration of electromagnetic energy, and easily gate tunable. These experiments were not optimized to minimize DP losses. Microscopic calculations~\cite{principi_prb_2013,principi_prbr_2013} indicate that these can be strongly reduced by using hBN (rather than e.g.~${\rm SiO}_2$) as a substrate.
\begin{figure}
\includegraphics[width=\linewidth]{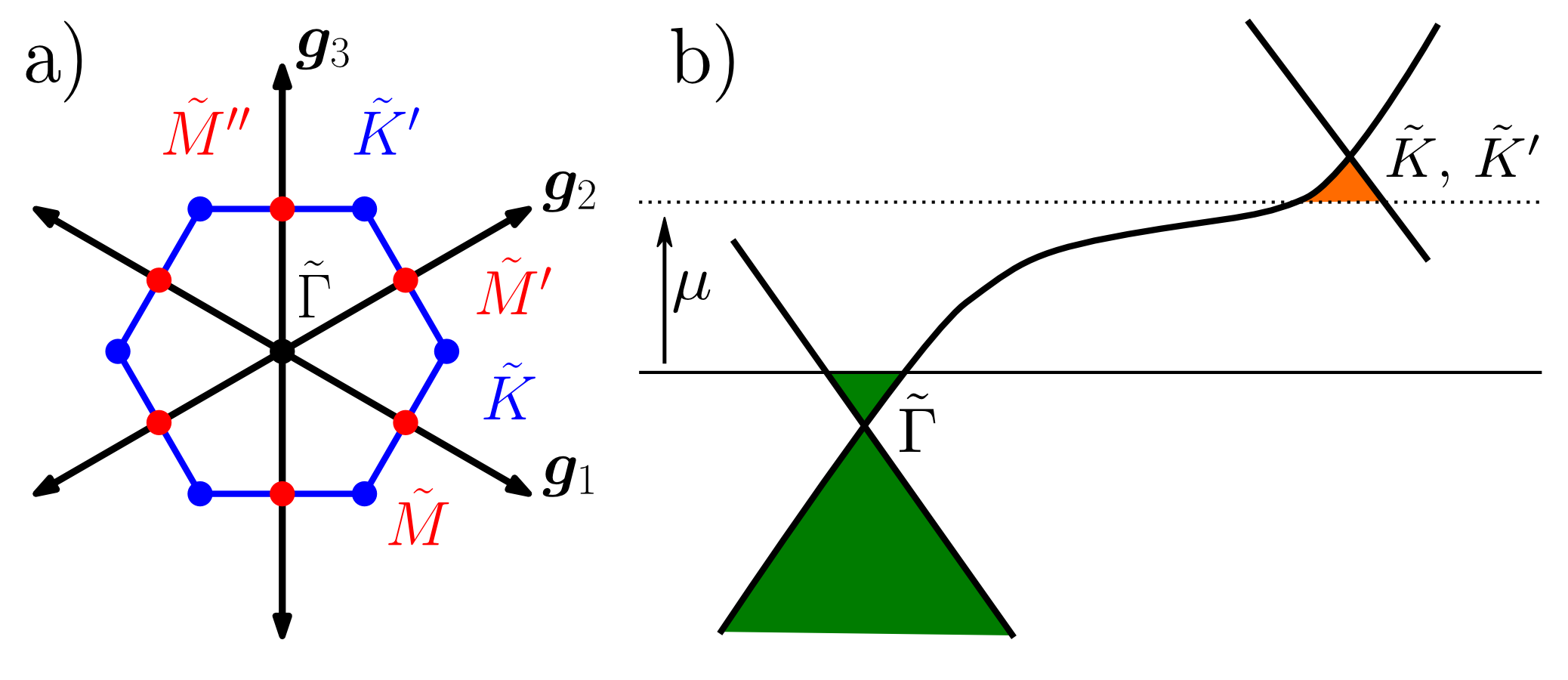}
\caption{\label{fig:cartoon} (Color online) Panel a) shows the SBZ with its high-symmetry points and the first star of reciprocal lattice vectors $\pm {\bm g}_1$, $\pm {\bm g}_2$, and $\pm {\bm g}_3$.
Here, the $\tilde{\Gamma}$ point (i.e.~the center of the SBZ) coincides with the $\nu = K, K^\prime$ points of the original graphene's Brillouine zone.
Panel b) illustrates the concept of plasmon morphing. Two Dirac-band crossings are located at different points of the SBZ and are separated in energy.
When the chemical potential $\mu$ lies at the position of the horizontal solid line, the system displays the plasmon~\cite{grapheneplasmonics,Diracplasmons} of a 2D Dirac system with $n$-type doping (green shaded area).
Upon increasing the chemical potential (dashed line), the $\tilde{\Gamma}$-point plasmon morphs into the $\tilde{K}$/$\tilde{K}^\prime$-point plasmon of a 2D Dirac system with $p$-type doping (orange shaded area).}
\end{figure}

In this work we try and combine ideas of these two fields of research by presenting a theory of the impact of a moir\'e superlattice on graphene's plasmons. Following recent transport experiments~\cite{ponomarenko_nature_2013,dean_nature_2013,hunt_science_2013}, we focus our attention on the case of long-wavelength superlattices since, in this case, interesting features in the miniband structure appear at carrier densities that can be achieved by standard electrostatic gating. By employing linear response theory within the random phase approximation~\cite{Giuliani_and_Vignale}, we calculate the plasmon modes of 2D MDFs in a moir\'e superlattice. We have found that graphene/hBN superlattices harbor a wealth of satellite plasmons.
These modes emerge from electron/hole pockets located at the high-symmetry, e.g.~$\tilde{M}$, $\tilde{K}$, and $\tilde{K}^\prime$, points in the SBZ---see Fig.~\ref{fig:cartoon}a).
Depending on the nature of the moir\'e superlattice, the $\tilde{\Gamma}$-point plasmon, which at low doping is an ordinary DP mode~\cite{grapheneplasmonics}, 
may survive or {\it morph} into a satellite plasmon when the chemical potential increases. The concept of plasmon morphing is sketched in Fig.~\ref{fig:cartoon}b).
Spectroscopy of plasmons in graphene superlattices therefore reveals precious information on the properties of the miniband structure. Even more interestingly, graphene/hBN stacks offer a low-loss plasmonic platform where knobs other than electrostatic gating~\cite{grapheneplasmonics}---such as the twist angle between the graphene and hBN crystals---can be used for the manipulation of plasmons.

\section{Moir\'e superlattice minibands} 

We describe the problem of a MDF moving in a moir\'e superlattice with the following single-particle continuum-model Hamiltonian~\cite{wallbank_prb_2013}
\begin{equation}\label{eq:singleparticleHamiltonian}
{\cal H}_0 = \hbar v_{\rm F}{\bm \sigma} \cdot {\bm p}\tau_0 + \hbar v_{\rm F} {\bm \sigma}\cdot {\bm A}({\bm r})\tau_3 + V({\bm r})\sigma_0\tau_0 +\Delta({\bm r}) \sigma_3\tau_3
\end{equation}
acting on the four-component pseudospinor $(\Psi_{A, K},\Psi_{B, K},\Psi_{B, K^\prime},-\Psi_{A, K^\prime})^{\rm T}$. 
In Eq.~(\ref{eq:singleparticleHamiltonian}), ${\bm p} = - i \hbar \nabla_{\bm r}$ is the 2D momentum measured from the centers of the two graphene's principal valleys $\nu = K, K^\prime$, $v_{\rm F} \sim 10^6~{\rm m}/{\rm s}$ is the Fermi velocity in an isolated graphene sheet, and $\sigma_a, \tau_a$ with $a=0,1,2,3$ are ordinary $2\times2$ Pauli matrices acting on graphene's sublattice and principal-valley degrees-of-freedom, respectively ($\sigma_0$ and $\tau_0$ are identity matrices). Finally, $V({\bm r})$, ${\bm A}({\bm r})$, and $\Delta({\bm r})$ are potentials due to the moir\'e superlattice. Note that the terms containing ${\bm A}$ and $\Delta$ have different signs in the two principal valleys. Since $V({\bm r})$, ${\bm A}({\bm r})$, and $\Delta({\bm r})$ are periodic perturbations, we can expand them in the following form: $V({\bm r}) = \sum_{\bm G}V({\bm G})\exp{(i {\bm G}\cdot {\bm r})}$, ${\bm A}({\bm r}) = \sum_{\bm G}{\bm A}({\bm G})\exp{(i {\bm G}\cdot {\bm r})}$, and $\Delta({\bm r}) = \sum_{\bm G}\Delta({\bm G})\exp{(i {\bm G}\cdot {\bm r})}$, where ${\bm G}$ denotes the reciprocal lattice vectors (RLVs) of the moir\'e superlattice. Since
${\cal H}_0$ is block-diagonal in valley space, one can find the spectrum of ${\cal H}_0$ in each principal valley $\nu = K, K^\prime$, separately.
\begin{figure}
\begin{overpic}[height=2.1in]{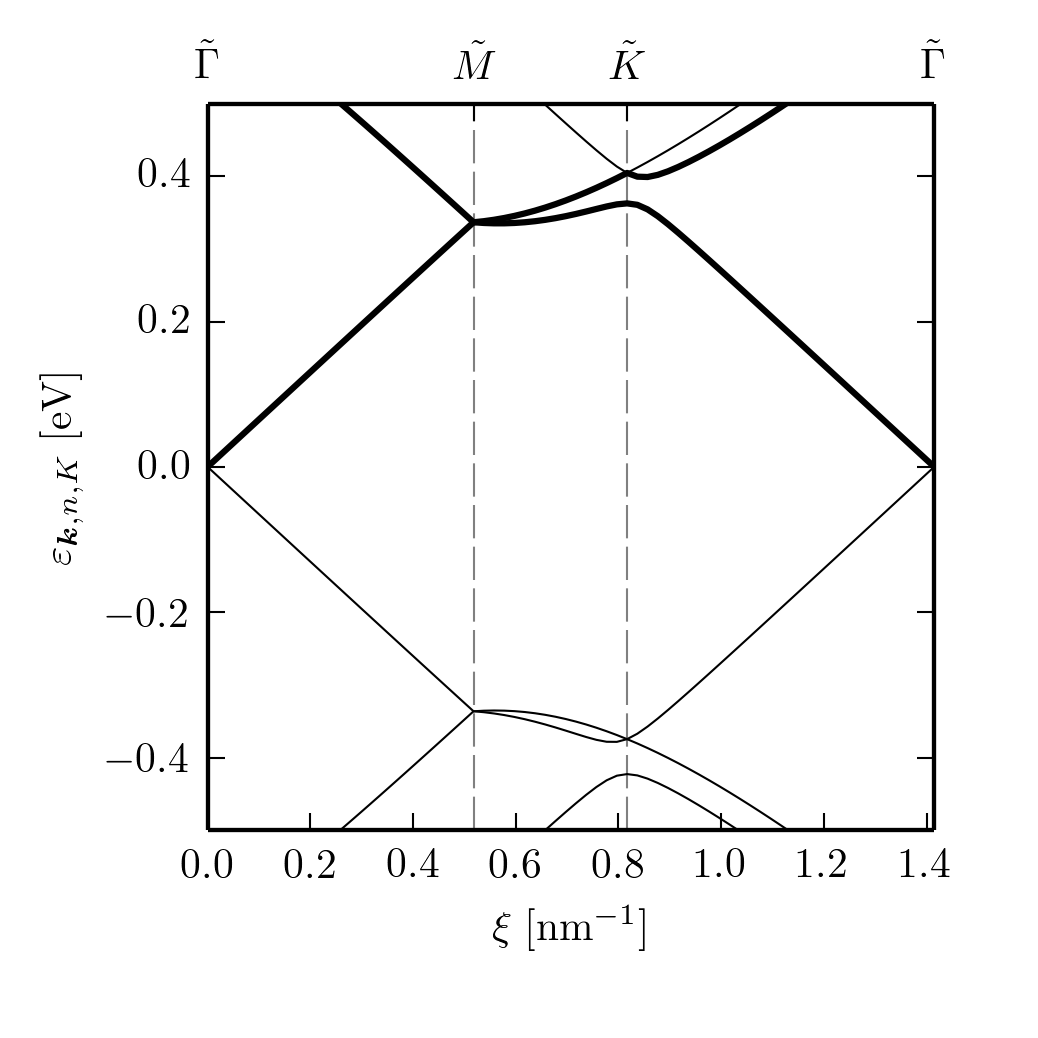}\put(2,90){a)}\end{overpic}
\begin{overpic}[height=2.1in]{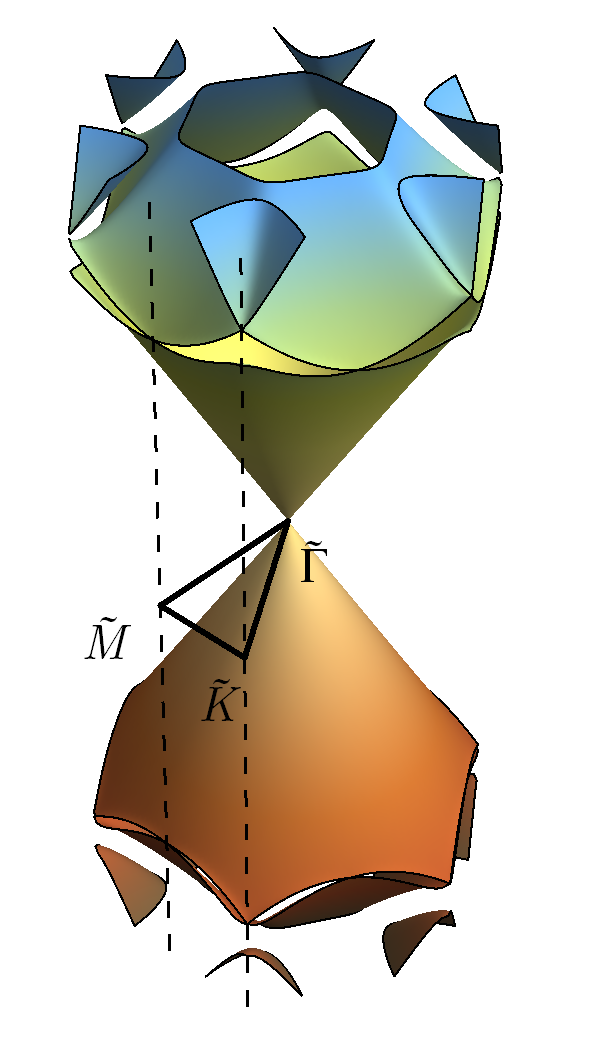}\put(2,90){b)}\end{overpic}
\begin{overpic}[height=2.1in]{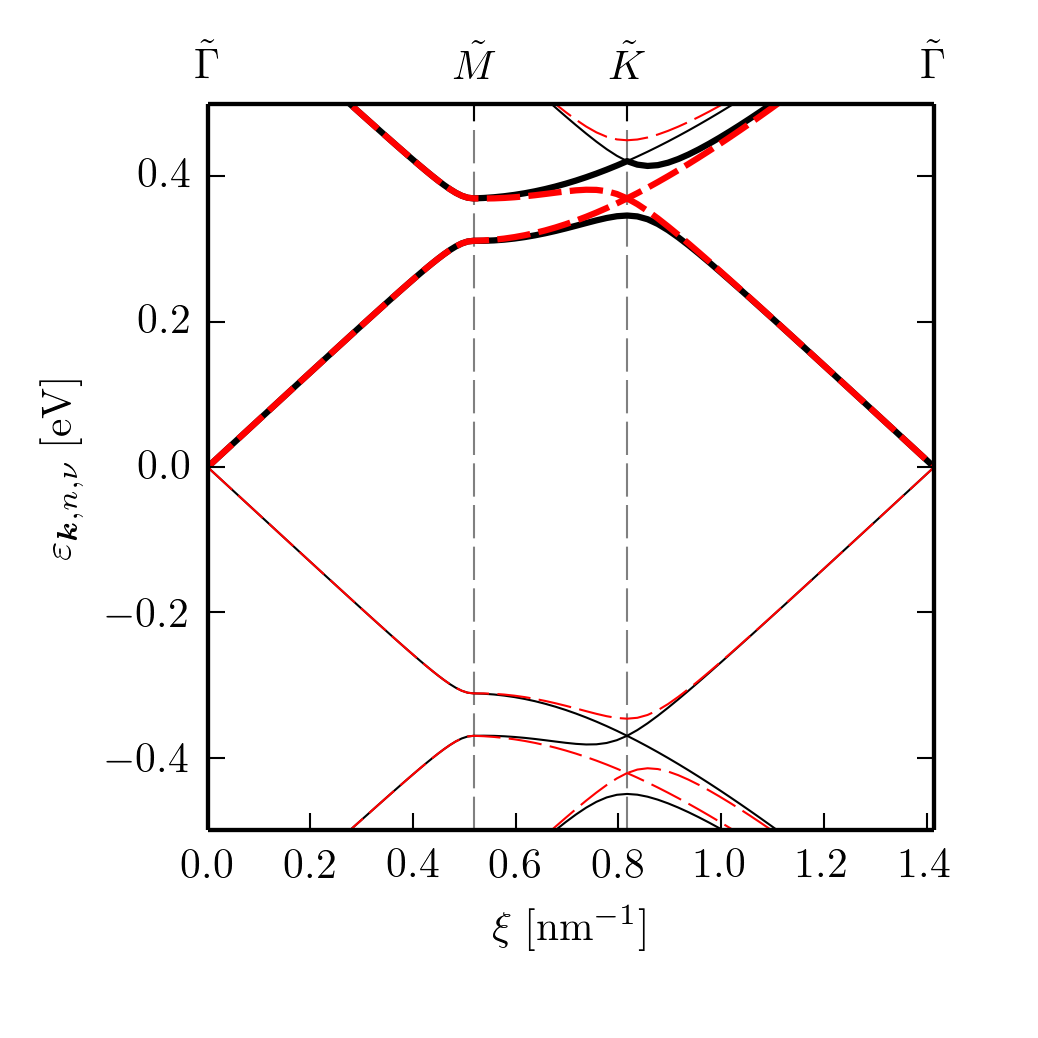}\put(2,90){c)}\end{overpic}
\begin{overpic}[height=2.1in]{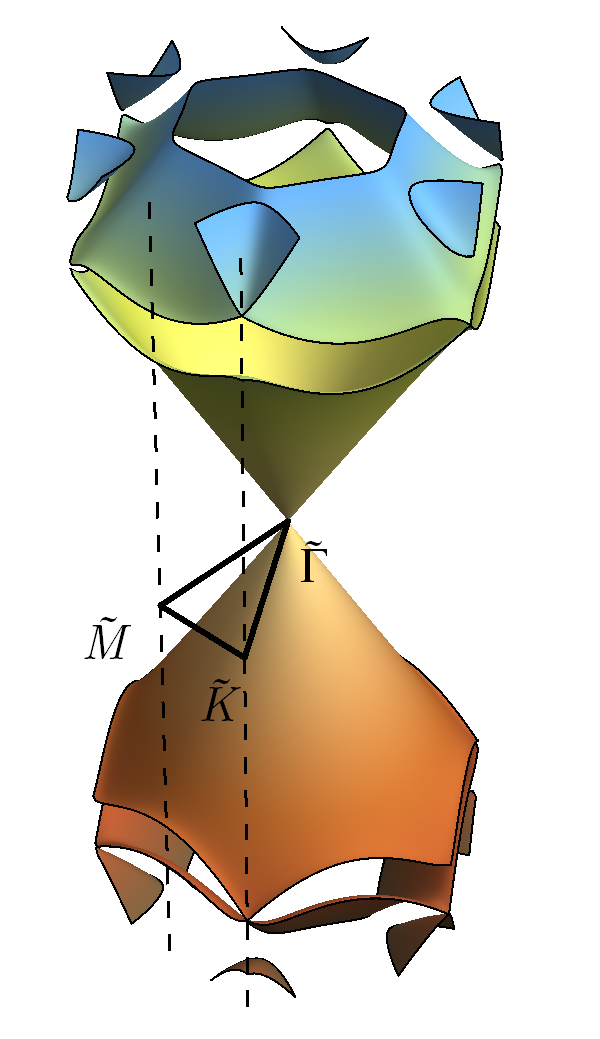}\put(2,90){d)}\end{overpic}
\caption{\label{fig:bands} (Color online) Graphene/hBN superlattice minibands along the $\tilde{\Gamma}$-$\tilde{M}$-$\tilde{K}$-$\tilde{\Gamma}$ direction (twist angle $\phi \simeq 0.03~{\rm rad}$, $\lambda \simeq 7~{\rm nm}$). Thick lines denote the first two conduction minibands. All the numerical results in this figure have been obtained by including ${\cal N} = 37$ RLVs ($|{\bm G}| \le 3 \times |{\bm g}_{1}|$) in Eq.~(\ref{eq:planewaveexpansion}). Panel a) refers to a moir\'e potential with
$V_{\rm s}= 30~{\rm meV}$ and $V_{\Delta} = 0$. Panel b) gives a 3D representation of the moir\'e superlattice miniband structure for the same parameters as in panel a). Panels c) and d) refer to a moir\'e potential with $V_{\rm s}= 0$ and $V_{\Delta} = 30~{\rm meV}$. In panel c) black solid  (red dashed) lines label the miniband structure around the $K$ ($K^\prime$) principal valley.}
\end{figure}

The eigenvectors $| {\bm k},n, \nu \rangle$ of ${\cal H}_0$ can be expanded in a plane-wave basis as
\begin{equation}\label{eq:planewaveexpansion}
\langle {\bm r} | {\bm k},n, \nu \rangle =  \frac{1}{\sqrt{L^{2}}}  \sum_{\bm G} {\bm u}_{\bm G}({\bm k},n, \nu)e^{i({\bm k} + {\bm G}) \cdot {\bm r}}~,
\end{equation}
where $L^2$ is the 2D electron system area and $\sum_{\bm G} {\bm u}_{\bm G}({\bm k},n', \nu')^{\dag} {\bm u}_{\bm G}({\bm k},n, \nu) = \delta_{n,n'}\delta_{\nu,\nu'}$. The wave vector ${\bm k}$ varies in the first SBZ, $n$ is a discrete index for the superlattice minibands, and $\nu = K, K^\prime$. Eigenvalues $\varepsilon_{{\bm k}, n, \nu}$ and eigenvector components ${\bm u}_{\bm G}({\bm k},n, \nu)$ can be found by solving the secular equation~\cite{Ashcroft_and_Mermin}
\begin{equation}\label{eq:secular_equation}
\sum_{{\bm G}'} {\cal H}_{{\bm G}, {\bm G}'}({\bm k}){\bm u}_{{\bm G}'}({\bm k},n, \nu) =\varepsilon_{{\bm k},n, \nu} {\bm u}_{{\bm G}}({\bm k},n,\nu)~,
\end{equation}
where ${\cal H}_{{\bm G}, {\bm G}'}({\bm k}) \equiv \hbar v_{\rm F} {\bm \sigma} \cdot ({\bm k} + {\bm G})\delta_{{\bm G},{\bm G}'}\tau_0  + \hbar v_{\rm F}{\bm \sigma}\cdot {\bm A}({\bm G}- {\bm G}')\tau_3 + V({\bm G}-{\bm G}')\sigma_0 \tau_0 + \Delta({\bm G}- {\bm G}')\sigma_3 \tau_3$. The size of the matrix ${\cal H}_{{\bm G}, {\bm G}'}({\bm k})$ is $4{\cal N} \times 4{\cal N}$, where ${\cal N}$ is the number of RLVs included in the expansion in Eq.~(\ref{eq:planewaveexpansion}).

The sparseness of ${\cal H}_{{\bm G}, {\bm G}'}({\bm k})$ is controlled by the number of RLVs included in the Fourier representation of the moir\'e potentials $V$, ${\bm A}$, and $\Delta$. Following Ref.~\onlinecite{graphenesuperlattices,wallbank_prb_2013}, we use only six RLVs $\pm {\bm g}_1$, $\pm {\bm g}_2$, and $\pm {\bm g}_3$ in the Fourier expansion of the moir\'e potentials. For two twisted honeycomb lattices with lattice constants $a$ and $a(1+\delta)$, the RLV ${\bm g}_1$ is given by~\cite{yankowitz_naturephys_2012} ${\bm g}_1 = 4\pi(\cos(\theta),\sin(\theta))/(\sqrt{3}\lambda)$, where $\lambda = (1+\delta)a/\sqrt{2(1+\delta)[1-\cos(\phi)] + \delta^2}$ is the moir\'e superlattice wavelength and $\phi$ the relative rotation angle between the two honeycomb lattices. Finally, $\theta$ is the relative rotation angle of the moir\'e pattern with respect to the graphene lattice with $\tan(\theta) = \sin(\phi)/[(1+\delta) - \cos(\phi)]$. The RLV ${\bm g}_2$ (${\bm g}_3$) can be obtained from ${\bm g}_1$ by a counterclockwise rotation of $\pi/3$ ($2\pi/3$), as shown in Fig.~\ref{fig:cartoon}a).
For graphene on hBN, $a = \sqrt{3}a_0$ is the graphene lattice constant ($a_0 \simeq 1.42~{\rm \AA}$) and $\delta \simeq 1.8\%$.

For the sake of simplicity, we focus our attention on the case in which pseudomagnetic fields are absent, i.e.~${\bm A}({\bm r}) = {\bm 0}$, although the theoretical apparatus described below is completely general. We then write the scalar potential as $V({\bm r}) = 2 V_{\rm s}\sum_{m =1\dots3}\cos{({\bm g}_m \cdot {\bm r})}$ and the mass term as $\Delta({\bm r})  = -2 V_{\Delta}\sum_{m=1\dots3} \sin{({\bm g}_{m} \cdot {\bm r})}$, implying that the Hamiltonian ${\cal H}_0$ is inversion-symmetric~\cite{footnote_asymmetry}. Illustrative numerical results for the minibands of a long-wavelength ($\phi \ll 1$) graphene/hBN moir\'e superlattice are reported in Fig.~\ref{fig:bands}. We focus our attention on features of the miniband structure occuring in the first two conduction minibands---thick lines in Fig.~\ref{fig:bands}a) and c).  Features at these energies can be accessed via electrostatic doping.

For $V_{\rm s} \neq 0$ and $V_{\Delta} =0$, Fig.~\ref{fig:bands}a), we see that the spectrum hosts a crossing
at the $\tilde{M}$ point of the SBZ.
This crossing evolves into an isolated Dirac point for scalar potentials with larger amplitude, e.g.~$V_{\rm s} \sim 100~{\rm meV}$. Since experiments~\cite{yankowitz_naturephys_2012,ponomarenko_nature_2013,dean_nature_2013,hunt_science_2013} seem to indicate weaker potentials, we have decided to use $V_{\rm s}  = 30~{\rm meV}$. Moreover, plasmons emerging from an isolated Dirac point have well studied properties~\cite{grapheneplasmonics}. For $V_{\rm s} = 0$ and $V_{\Delta} \neq 0$, Fig.~\ref{fig:bands}b), the spectrum shows satellite Dirac points at the $\tilde{K}$ (in one principal valley) and $\tilde{K}^\prime$ (in the other principal valley) points of the SBZ. See Appendix~\ref{app:technicalities} for further important considerations on Fig.~\ref{fig:bands}c).

\section{Plasmons in a moir\'e superlattice} 

Complete information on the plasmon modes of an {\it interacting} system of MDFs in a moir\'e superlattice is contained in the density response function~\cite{Giuliani_and_Vignale} $\chi_{{\bm G}, {\bm G}'}({\bm q}, \omega)  \equiv
\chi_{nn}({\bm q} +{\bm G}, {\bm q} + {\bm G}', \omega)$, viewed as a matrix with respect to the RLVs ${\bm G}, {\bm G}'$ and with ${\bm q}$ spanning the first SBZ. We also introduce the inverse dielectric matrix~\cite{Giuliani_and_Vignale}
\begin{equation}\label{eq:inversedielectricmatrix}
[\epsilon^{-1}]_{{\bm G},{\bm G}'}({\bm q},\omega) = \delta_{{\bm G},{\bm G}'} + v_{{\bm G}}({\bm q}) \chi_{{\bm G},{\bm G}'}({\bm q},\omega)~,
\end{equation}
where $v_{{\bm G}}({\bm q}) = v({\bm q} + {\bm G})$ with $v(q) = 2\pi e^2/(\epsilon q)$ the 2D Fourier transform of the Coulomb potential. Here $\epsilon  = (\epsilon_1 + \epsilon_2)/2$ is the average of the dielectric constants of the media above ($\epsilon_1$) and below ($\epsilon_2$) the graphene flake. For graphene with one side exposed to air and one to hBN, $\epsilon_1 =1$ and $\epsilon_2 \simeq 4.5$. The value of $\epsilon_2$ has been taken from Ref.~\onlinecite{yu_pnas_2013}.

A good starting point to calculate plasmons in electron liquids is the so-called random phase approximation~\cite{Giuliani_and_Vignale} (RPA) in which the full density-density response function $\chi_{{\bm G},{\bm G}'}({\bm q},\omega)$ in Eq.~(\ref{eq:inversedielectricmatrix}) is approximated by the solution of the following Dyson's equation:
\begin{eqnarray}\label{eq:crystalRPA}
\chi_{{\bm G},{\bm G}'}({\bm q},\omega) &=& \chi^{(0)}_{{\bm G},{\bm G}'}({\bm q},\omega) \nonumber\\
&+& \sum_{{\bm G}''}
\chi^{(0)}_{{\bm G},{\bm G}''}({\bm q},\omega)v_{{\bm G}''}({\bm q})\chi_{{\bm G}'',{\bm G}'}({\bm q},\omega)
\end{eqnarray}
where $\chi^{(0)}_{{\bm G},{\bm G}'}({\bm q},\omega)$ is the density response function of the
{\it non-interacting} electron system in the moir\'e superlattice.
Off-diagonal terms with respect to RLVs in Eq.~(\ref{eq:crystalRPA}) represent crystal local field effects, which we retain since they may be important
in comparing theory with experimental results~\cite{alkauskas_prb_2013}.

The quantity $\chi^{(0)}_{{\bm G},{\bm G}'}({\bm q},\omega)$ is given by the following expression:
\begin{equation}
\begin{split}\label{eq:moireLindhard}
&\chi^{(0)}_{{\bm G},{\bm G}'}({\bm q},\omega) =  \frac{2}{L^{2}}\sum_{{\bm k},n; {\bm k}', n'; \nu} \frac{n_{\rm F}(\varepsilon_{{\bm k},n,\nu}) - n_{\rm F}(\varepsilon_{{\bm k}',n',\nu})}{\hbar \omega + \varepsilon_{{\bm k},n,\nu} - \varepsilon_{{\bm k}',n',\nu} + i \eta}  \\
&\times  {\cal M}_{{\bm k}, n, \nu; {\bm k}', n', \nu}({\bm q} + {\bm G}) {\cal M}^\dagger_{{\bm k}, n, \nu; {\bm k}', n', \nu}({\bm q} + {\bm G}')
\end{split}
\end{equation}
where the factor two accounts for spin degeneracy, $\eta$ is a positive infinitesimal, $n_{\rm F}(x)  = \{\exp[(x- \mu)/k_{\rm B} T] +1\}^{-1}$ is the Fermi-Dirac occupation factor at temperature $T$ and chemical potential $\mu$. Finally, ${\cal M}_{{\bm k}, n, \nu; {\bm k}', n', \nu}({\bm q} + {\bm G}) \equiv \langle {\bm k},n,\nu | e^{-i({\bm q} + {\bm G}) \cdot {\bm r}} | {\bm k}',n',\nu\rangle$. We emphasize that Eq.~(\ref{eq:moireLindhard}) is the sum of two contributions, one for each principal valley $\nu = K, K^\prime$.

Self-sustained oscillations of an electron system in a crystal can be found~\cite{Giuliani_and_Vignale} by solving the equation ${\rm det}\{[1/\chi]_{{\bm G},{\bm G}'}({\bm q},\omega)\} = 0$.  Alternatively, one can directly calculate the {\it loss function} $L({\bm q},\omega) \equiv - \Im m\{[1/\epsilon]_{{\bm 0}, {\bm 0}}({\bm q},\omega)\}$, which is appealing since it is directly measured by electron-energy-loss spectroscopy~\cite{egerton_rpp_2009}. The loss function displays sharp peaks at the plasmon poles and carries also precious information on inter-band transitions and Landau damping. The latter determines the width of the plasmon peak in $L({\bm q}, \omega)$. In this work we focus our attention on $L({\bm q}, \omega)$.

\section{Numerical results and discussion}

\begin{figure}
\includegraphics[width=\linewidth]{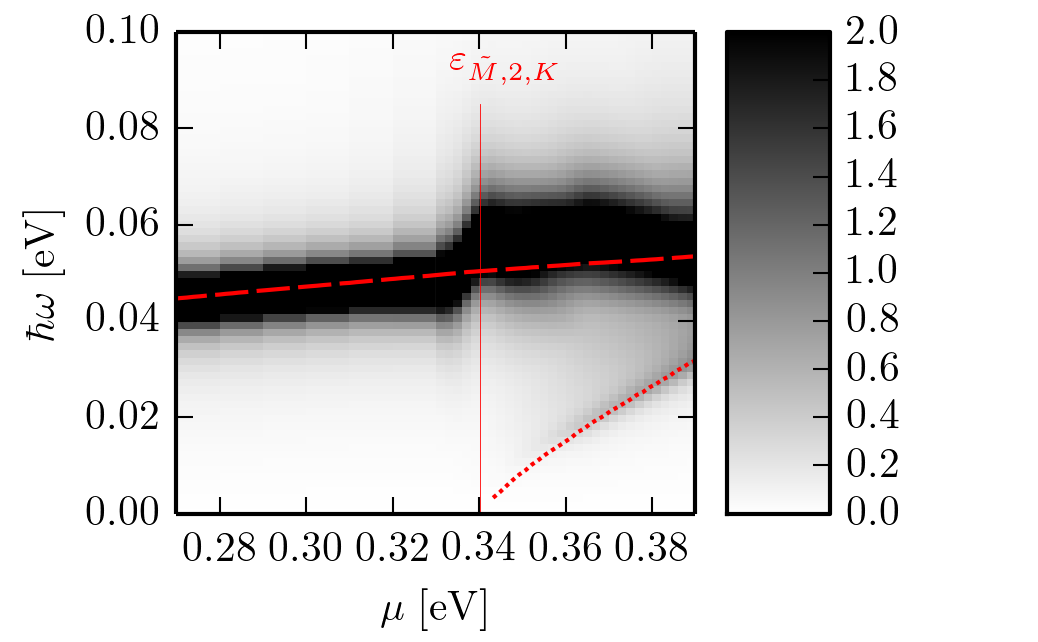}
\caption{\label{fig:scalarpotential} (Color online) A 2D density plot of the RPA loss function $L({\bm q}, \omega)$ for the superlattice miniband structure in Fig.~\ref{fig:bands}a) and $|{\bm q}| = 0.007~{\rm nm}^{-1}$.
The vertical thin solid line denotes the bottom edge $\varepsilon_{\tilde{M}, 2, K}$ of the $n=2$ conduction miniband---see Fig.~\ref{fig:bands}a). The long-dashed line represents the chemical potential dependence of a DP.  For chemical potentials above $\varepsilon_{\tilde{M}, 2, K}$, a satellite $\tilde{M}$-point plasmon is generated: its chemical potential dependence (dotted line) follows that of a plasmon in a 2D parabolic-band electron gas. 1D cuts of this 2D density plot for different values of the chemical potential $\mu$ are reported in Appendix~\ref{app:moreresults}.}
\end{figure}

\begin{figure}
\begin{overpic}[width=\linewidth]{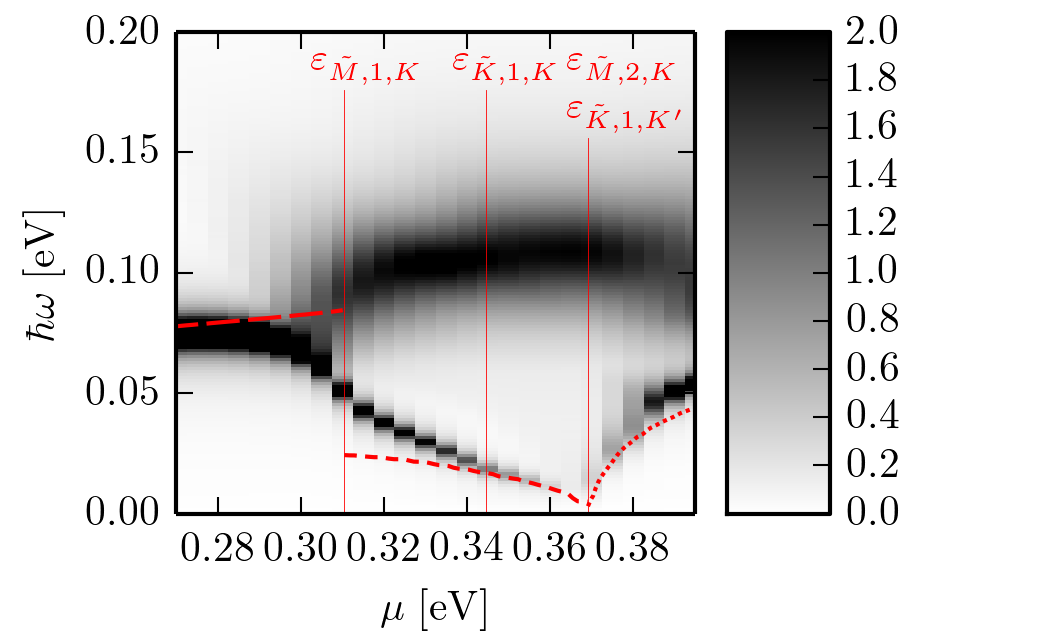}\put(2,90){}\end{overpic}
\caption{(Color online)
Same as in Fig.~\ref{fig:scalarpotential} but for the miniband structure shown in Fig.~\ref{fig:bands}c) and $|{\bm q}| = 0.021~{\rm nm}^{-1}$. The long-dashed and dotted lines have the same physical meaning as in Fig.~\ref{fig:scalarpotential}. The short-dashed line represents the chemical potential dependence
of a $\tilde{K}$-point plasmon stemming from a satellite Dirac point. Note that $\varepsilon_{\tilde{K},1,K'}$ coincides with $\varepsilon_{\tilde{M},2,K}$.\label{fig:masspotential}}
\end{figure}

\begin{figure}
\begin{overpic}[width=\linewidth]{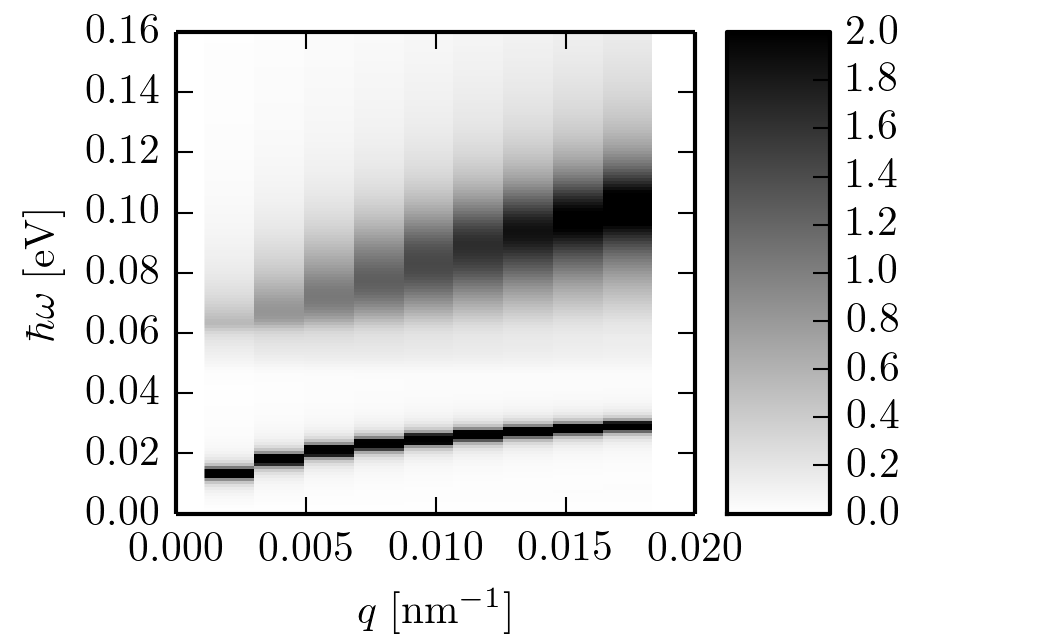}\put(2,55){a)}\end{overpic}
\begin{overpic}[width=\linewidth]{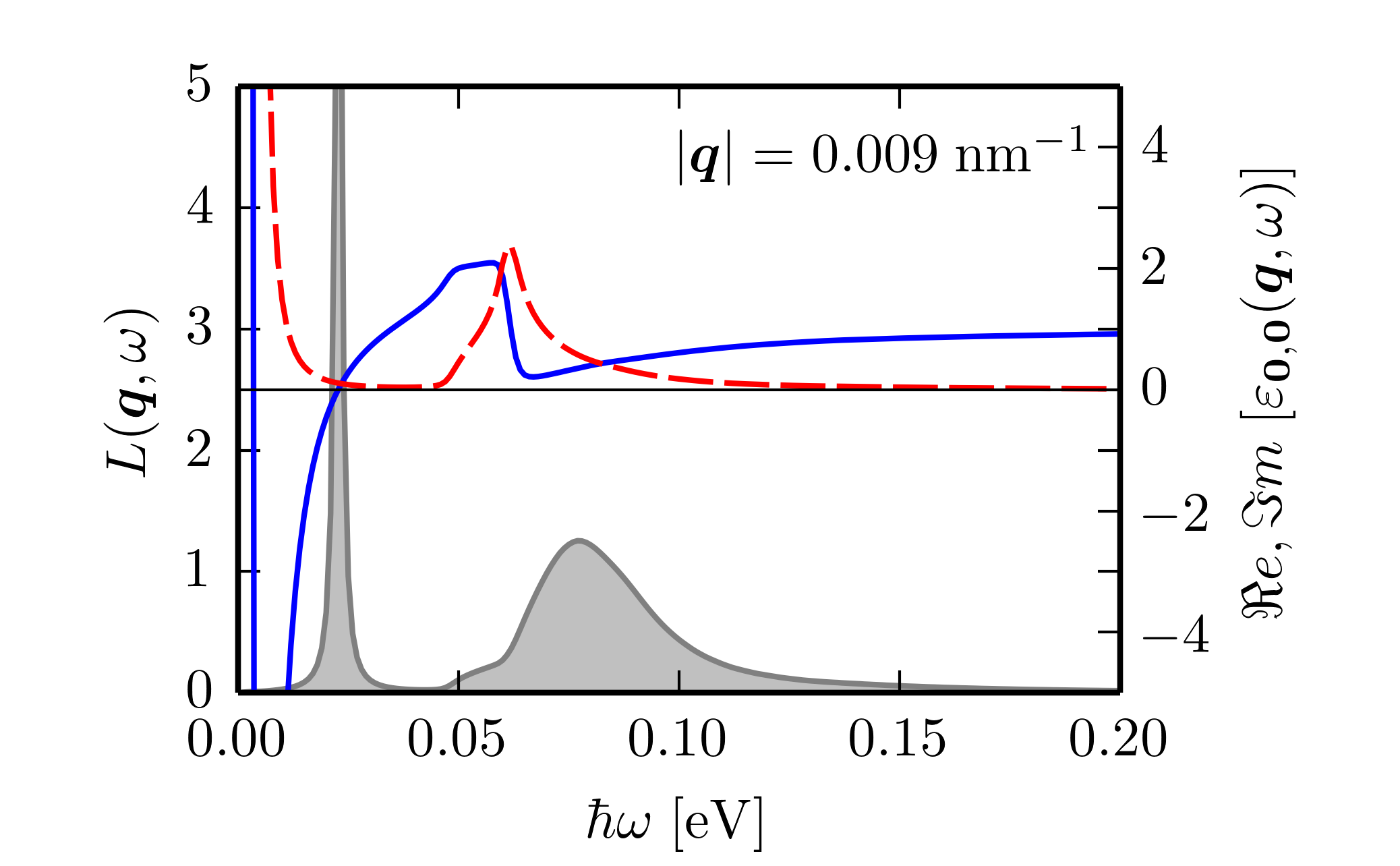}\put(2,55){b)}\end{overpic}
\caption{\label{fig:qdependence} (Color online) Wave vector and energy dependence of the
loss function $L({\bm q}, \omega)$ for $\mu = 330~{\rm meV}$.
All the other parameters are as in Fig.~\ref{fig:masspotential}. In this plot $L({\bm q}, \omega)$ has been evaluated for $9$ values of $q$. The grey-shaded area in panel b) illustrates the energy dependence of $L({\bm q}, \omega)$ for $q = 0.009~{\rm nm}^{-1}$. Solid and dashed curves represent $\Re e[\epsilon_{{\bm 0}, {\bm 0}}({\bm q}, \omega)]$ and $\Im m[\epsilon_{{\bm 0}, {\bm 0}}({\bm q}, \omega)]$, respectively. The values of these functions can be inferred from the vertical axis on the right. While the sharp peak at $\hbar\omega \simeq 20~{\rm meV}$ corresponds to a true zero of the macroscopic dielectric function $\epsilon_{{\bm 0}, {\bm 0}}({\bm q}, \omega)$, the broad peak at $\hbar \omega\simeq 100~{\rm meV}$ does not.}
\end{figure}

A summary of our main results for the RPA loss function $L({\bm q}, \omega)$---calculated  at $T= 10~{\rm K}$ and for the illustrative moir\'e miniband structures in Figs.~\ref{fig:bands}a),b)---is reported in Figs.~\ref{fig:scalarpotential}-\ref{fig:qdependence}. All the results shown in this work have been obtained for a wave vector ${\bm q}$ oriented along
the $\tilde{\Gamma}$-$\tilde{M}$ direction. In the range of parameters explored in this work we have not noticed significant angular anisotropies of the satellite plasmons. (This has been checked by performing numerical calculations with ${\bm q}$ oriented along the $\tilde{\Gamma}$-$\tilde{K}$.) Technical details relative to the numerical approach have been reported in Appendix~\ref{app:technicalities}. Additional numerical results and discussions have been reported in Appendix~\ref{app:moreresults}.

In Fig.~\ref{fig:scalarpotential} we plot the loss function $L({\bm q}, \omega)$ for the superlattice miniband structure shown in Fig.~\ref{fig:bands}a). For chemical potentials below the bottom edge $\varepsilon_{\tilde{M}, 2, K}$ of the second conduction miniband, the loss function peaks at the usual DP mode, i.e. $\tilde{\Gamma}$-point plasmon (dashed line). For chemical potentials {\it above} $\varepsilon_{\tilde{M}, 2, K}$, we clearly see that a new satellite plasmon mode is generated at the $\tilde{M}$ point. Interestingly, the dotted line in Fig.~\ref{fig:scalarpotential}, which tracks the chemical potential dependence of the new $\tilde{M}$-point plasmon, corresponds to the analytical formula of a plasmon
excitation in a 2D parabolic-band electron gas~\cite{Giuliani_and_Vignale}, i.e.~$\omega^2_{\rm 2DEG} = q \times 2\pi n_{\tilde{M}, K}(\mu) e^2/(m^\star \epsilon)$, with band mass $m^\star \simeq 0.01~m_{\rm e}$, $m_{\rm e}$ being electron's mass in vacuum. Here, the quantity $n_{\kappa, \nu}(\mu)$ represents the density of a pocket of electrons or holes at the high-symmetry point $\kappa$ of the SBZ, in the valley $\nu = K, K^\prime$, and for a chemical potential $\mu$. The dependence of $n_{\kappa, \nu}$ on $\mu$ is discussed in Appendix~\ref{app:dependence} at $T=0$. The parabolic-band-like (i.e.~like $n^{1/2}_{\tilde{M}, K}$ rather than $n^{1/4}_{\tilde{M}, K}$, as expected for a DP~\cite{grapheneplasmonics,Diracplasmons}) density dependence of this satellite mode is attributed to the parabolic dependence of the superlattice minibands on ${\bm k}$
near $\varepsilon_{\tilde{M},2,K}$ and for ${\bm k}$ along the $\tilde{M}$-$\tilde{K}$ direction---see Fig.~\ref{fig:bands}a).

The situation is even richer in the case $V_{\rm s} = 0$ and $V_{\Delta} \neq 0$. Representative results are shown in Fig.~\ref{fig:masspotential}, where we show the loss function $L({\bm q}, \omega)$ for the miniband structure in Fig.~\ref{fig:bands}b). Increasing the chemical potential, the ordinary $\tilde{\Gamma}$-point DP morphs into a $\tilde{K}$-point DP with lower energy. Fig.~\ref{fig:qdependence}a) shows that this mode displays a 2D dispersion $\propto \sqrt{q}$. Moreover, its chemical potential dependence is consistent with that of DPs~\cite{grapheneplasmonics}, i.e.~$\omega^2_{\rm DP} =  q \times 2\sqrt{\pi}v_{{\rm F}, \tilde{K}, K'}  n^{1/2}_{\tilde{K}, K'}(\mu)e^2/(\hbar \epsilon)$ with an effective Fermi velocity $v_{{\rm F}, \tilde{K}, K'} \simeq 0.3~v_{\rm F}$, which is reduced with respect to the Fermi velocity $v_{\rm F}$ in an isolated graphene sheet. Satellite Dirac points in the superlattice miniband structure enable
$\tilde{K}$-point DP modes with low energy for dopings $\gtrsim 350~{\rm meV}$. Long-wavelength graphene superlattices give therefore access to long-lived low-energy, e.g. Terahertz, plasmons, which
are difficult to reach due to the ultralow dopings ($\simeq 10~{\rm meV}$) that these modes require in the absence of a superlattice~\cite{grapheneplasmonics,Diracplasmons} (ultralow carrier densities imply strong susceptibility to disorder and, in turn, short plasmon lifetimes).

A further increase in $\mu$ generates a satellite $\tilde{M}$-point plasmon similar to that in Fig.~\ref{fig:scalarpotential}, with the same effective mass $m^\star \simeq 0.01~m_{\rm e}$. In Fig.~\ref{fig:masspotential} we also notice a broad peak at energy $\hbar\omega \simeq 100~{\rm meV}$. As demonstrated in Fig.~\ref{fig:qdependence}b), this peak is due to inter-band electron-hole excitations, which are quite bunched in energy. Indeed, the real part of the macroscopic dielectric function $\epsilon_{{\bm 0}, {\bm 0}}({\bm q}, \omega)$ does not vanish for $\hbar\omega \simeq 100~{\rm meV}$ and $\Im m[\epsilon_{{\bm 0}, {\bm 0}}({\bm q}, \omega)]$ is large at the same energy.

Recently, it has been shown~\cite{song_arxiv_2014} that graphene on hBN can display topologically non-trivial bands (i.e.~bands yielding finite Chern numbers) in the case of commensurate stackings. It will be interesting to study the plasmonic properties of these special stacks, especially in a magnetic field. Due the superb electronic quality of graphene on hBN, the plasmon modes described above are characterized by very low damping rates~\cite{principi_prbr_2013,principi_prb_2013}. We truly hope that our predictions will stimulate scattering-type near-field optical~\cite{chen_nature_2012,fei_nature_2012} and electron-energy loss~\cite{egerton_rpp_2009} spectroscopy studies of the plasmonic properties of graphene/hBN stacks.

\acknowledgements It is a pleasure to thank Frank Koppens and Francesco Pellegrino for useful discussions. This work was supported by the European Community under Graphene Flagship (contract no. CNECT-ICT-604391), MIUR (Italy) through the programs ``FIRB - Futuro in Ricerca 2010" - Project PLASMOGRAPH (Grant No. RBFR10M5BT) and ``Progetto Premiale 2012'' - Project ABNANOTECH, MINECO (Spain) through Grant No. FIS2011-23713, and the European Research Council Advanced Grant (contract 290846). We have made use of free software (www.gnu.org, www.python.org).

\begin{figure}
\includegraphics[width=1.00\linewidth]{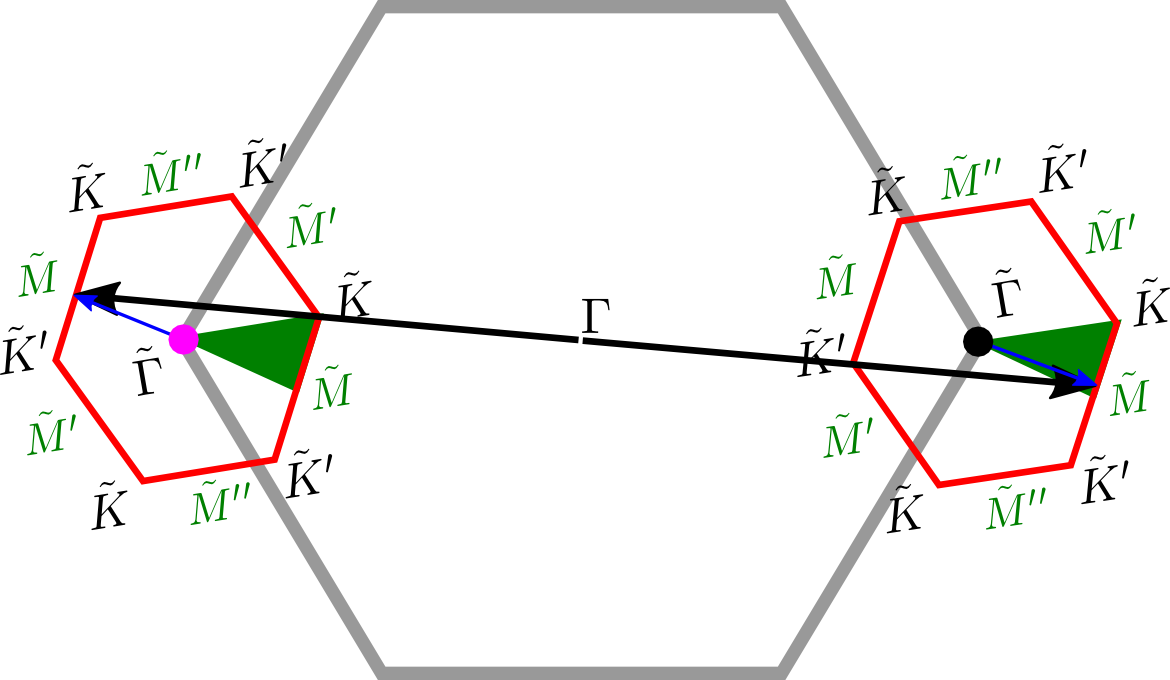}
\caption{(Color online) Schematic representation of the graphene's Brillouin zone (thick gray hexagon) and of two SBZs (red hexagons) around the graphene principal valleys $K$ (black filled circle) and $K'$ (magenta filled circle). The boundaries of the green-shaded triangular areas represent the path along which the miniband structure is shown in Fig.~\ref{fig:bands} in the main text.
The thick black arrows indicate two points in reciprocal space which are connected by the spatial inversion. The blue arrows indicate the same two points, with respect to the center $\tilde{\Gamma}$ of the SBZs. The two green triangles do not map onto each other under spatial inversion.\label{fig:sbzs}}
\end{figure}

\appendix

\section{Technical remarks}
\label{app:technicalities}

The results for the miniband dispersions in Fig.~\ref{fig:bands} in the main text have been obtained by using ${\cal N} = 37$, while those for the loss function in Figs.~\ref{fig:scalarpotential}-\ref{fig:qdependence} have been obtained by using ${\cal N} = 7$ (corresponding to the origin and the first ``star'' of RLVs). The infinitesimal parameter $\eta$ in Eq.~(\ref{eq:moireLindhard}) of the main text has been chosen to be mesh-dependent, defining it as the ratio between the width of the first conduction miniband and the total number of points ${\cal N}_{\rm SBZ}$ in the SBZ. All the results in this work have been obtained with ${\cal N}_{\rm SBZ}$ in the interval $22500 \leq {\cal N}_{\rm SBZ} \leq 230400$. 

Before discussing technical details on the calculation of the minibands, we  would like to make a comment on Fig.~\ref{fig:bands}c) in the main text. There we show the miniband structure along the path $\tilde{\Gamma}$-$\tilde{M}$-$\tilde{K}$-$\tilde{\Gamma}$ in the SBZ. In the case $V_{\rm s} = 0$ and $V_{\Delta} \neq 0$, the miniband dispersion is different in the two principal valleys of the graphene's Brillouin zone, i.e.~in the neighborhood of the $K$ and $K'$ points. At first sight, this difference may seem surprising because, as discussed in the main text, our Hamiltonian is symmetric under spatial inversion and the points $K$ and $K'$ map onto each other under this symmetry---see Fig.~\ref{fig:sbzs}.
However, we point out that the paths along which the minibands are shown do {\it not} map onto each other under space inversion. This is described in Fig.~\ref{fig:sbzs}. 
Indeed, the point $\tilde{K}$ in the $K$ valley is mapped onto the point $\tilde{K}'$ in the $K'$ valley, i.e.~{\it both} valleys and ``minivalleys'' are exchanged under spatial inversion.

\begin{figure}
\begin{overpic}[width=3.46in]{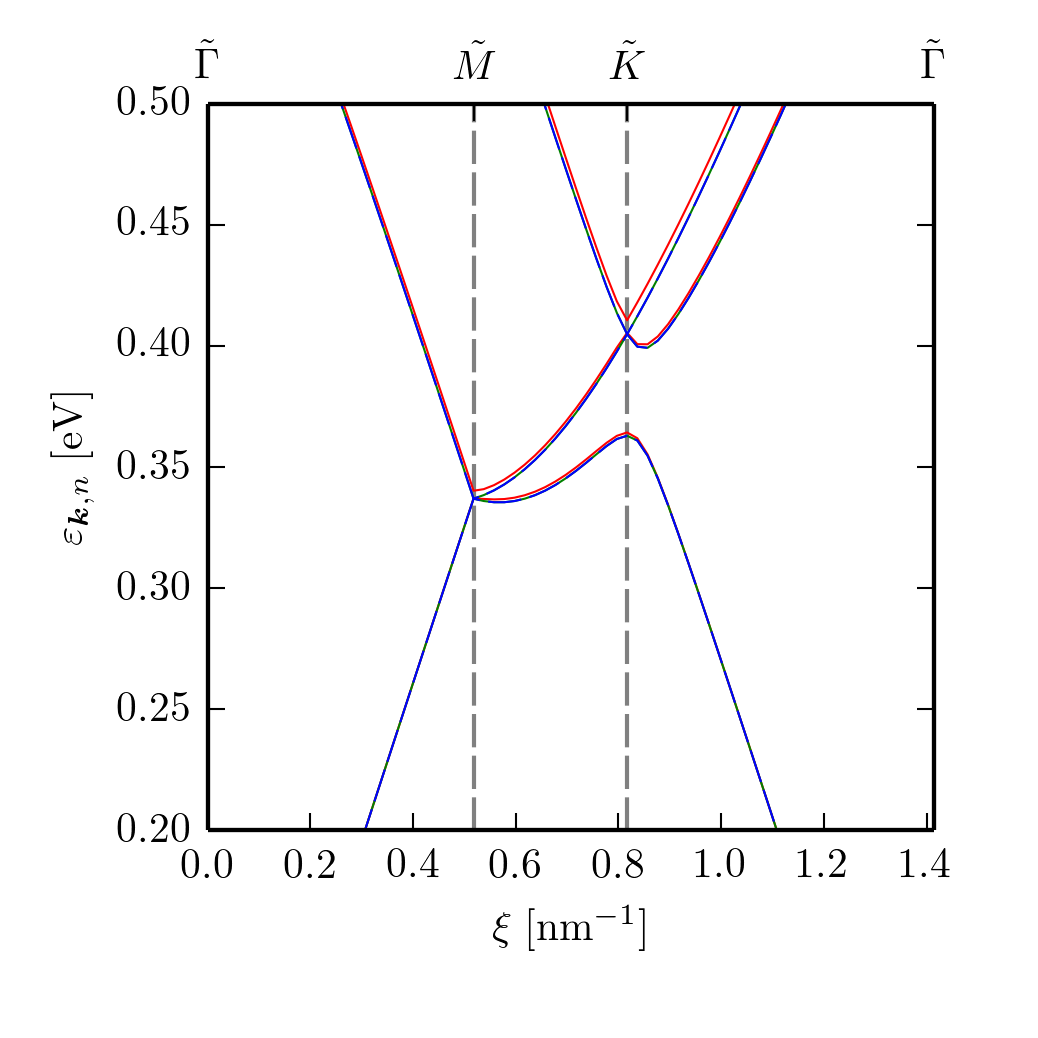}\put(2,90){a)}\end{overpic}
\begin{overpic}[width=3.46in]{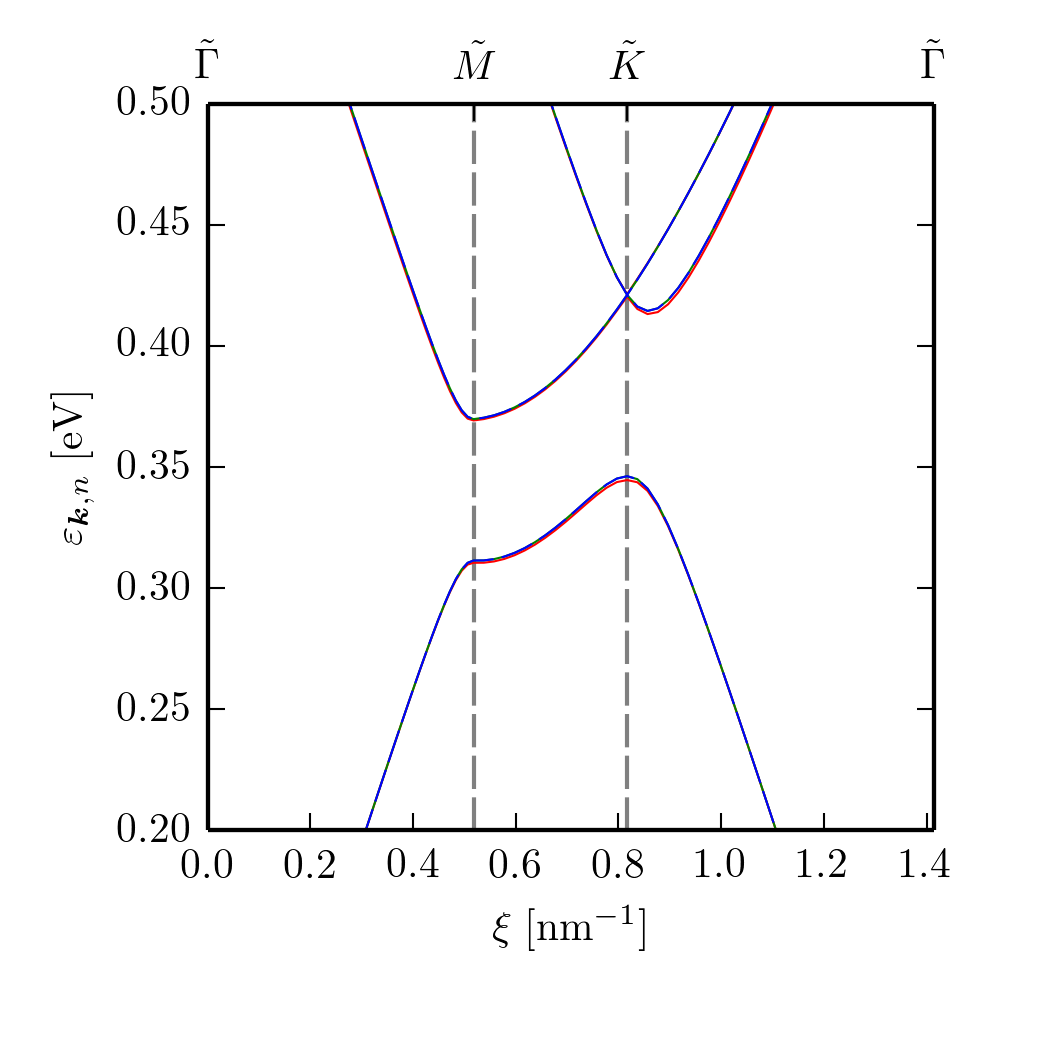}\put(2,90){b)}\end{overpic}
\caption{(Color online) Same as in Fig.~\ref{fig:bands} of the main text but for three different values of ${\cal N}$, i.e.~${\cal N} = 7$ (red), $19$ (green), and $37$ (blue), corresponding to $|{\bm G}| \le |{\bm g}_{1}|$, $2 \times |{\bm g}_{1}|$, and $3 \times |{\bm g}_{1}|$, respectively. All the other parameters in panels a) and b) of this Figure are as in panels a) and c) of Fig.~\ref{fig:bands} in the main text, respectively.  Panel b) of this Figure, however, shows data only for the $K$ principal valley. We can clearly see that calculations for different values of ${\cal N}$ are practically indistinguishable from each other, except for small deviations near the edges of the SBZ.\label{fig:bandsconvergence}}
\end{figure}

In Fig.~\ref{fig:bandsconvergence}, we show that, for the weak moir\'e potentials used in this work, the miniband dispersion does not change appreciably by increasing the number of RLVs. It is important to point out that the calculation of the loss function scales quadratically with the number of RLVs in the mesh [because of the double sum over all the bands in Eq.~(\ref{eq:moireLindhard}) of the main text]. An increase in the number ${\cal N}$ of RLVs for the calculation of the loss function is therefore computationally very expensive.
Moreover, the RLV mesh must respect the system symmetry: this implies that only the discrete set of values ${\cal N} = 7$, $19$, $37$, $\dots$ can be used.
Although we checked the accuracy of the data for the loss function by increasing ${\cal N}$ up to ${\cal N} = 19$ for a specific set of parameters, we found impractical to use ${\cal N} > 7$ when scans over the frequency $\omega$ and the chemical potential $\mu$/wave vector $q$ are needed.
\section{Additional numerical results}
\label{app:moreresults}

\begin{figure}[h!]
\begin{overpic}[width=3.46in]{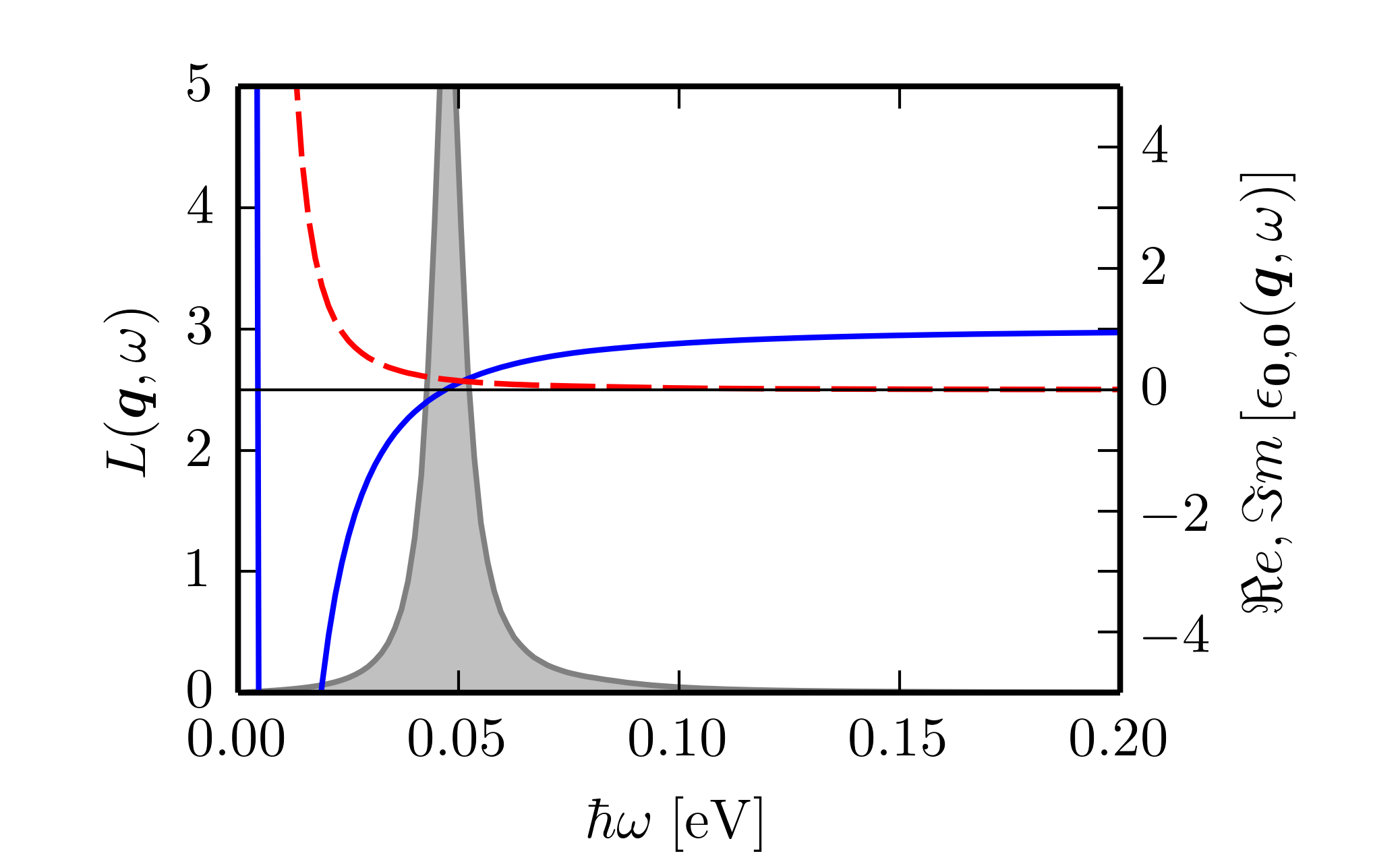}\put(2,60){a)}\end{overpic}
\begin{overpic}[width=3.46in]{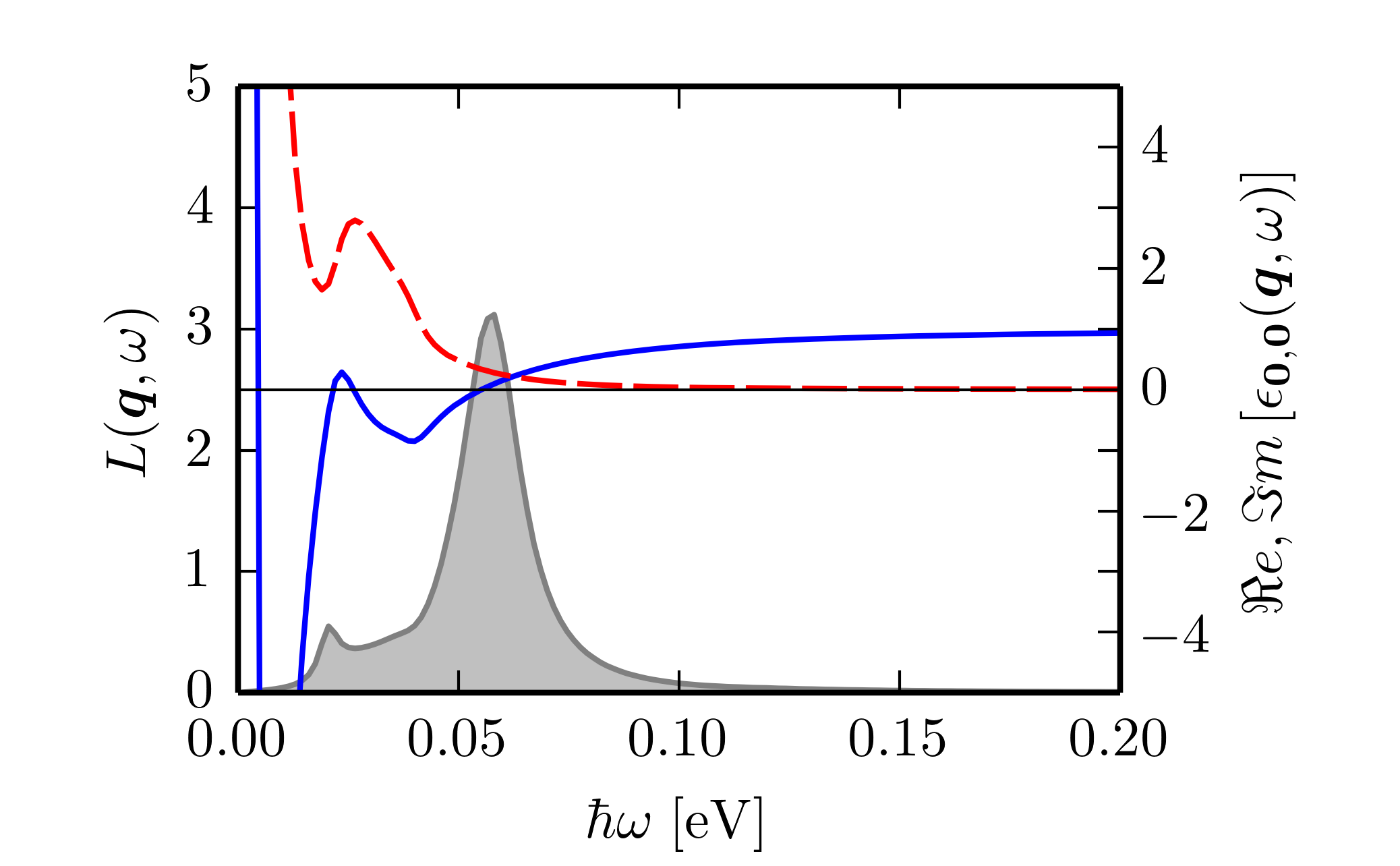}\put(2,60){b)}\end{overpic}
\caption{(Color online) Energy dependence of the loss function (grey-shaded area), calculated for a moir\'e potential as in Fig.~\ref{fig:bands}a) of the main text. The loss function is calculated at $|{\bm q}| = 0.007~{\rm nm}^{-1}$, as in Fig.~\ref{fig:scalarpotential} of the main text. Panel a) [b)] refers to a chemical potential $\mu = 0.330~{\rm eV}$ [$0.368~{\rm eV}$]. Solid and dashed lines refer to the real and imaginary part of the macroscopic dielectric function $\epsilon_{{\bm 0}, {\bm 0}}({\bm q}, \omega)$, respectively.
The values of these functions can be read from the vertical axis on the right. \label{fig:scalarpotential_appendix}}
\end{figure}
\begin{figure*}[h!]
\begin{overpic}[width=3.46in]{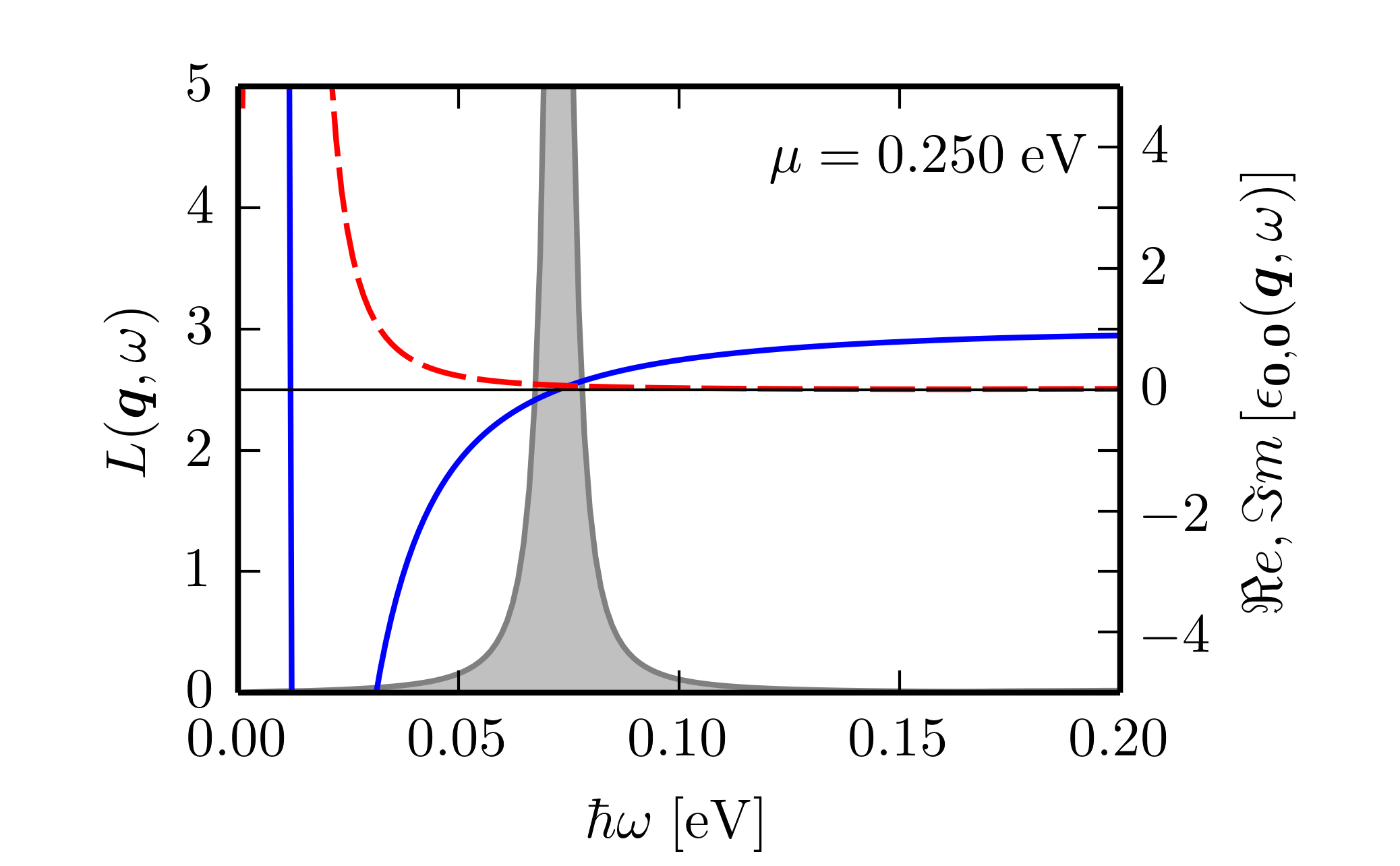}\put(2,60){a)}\end{overpic}
\begin{overpic}[width=3.46in]{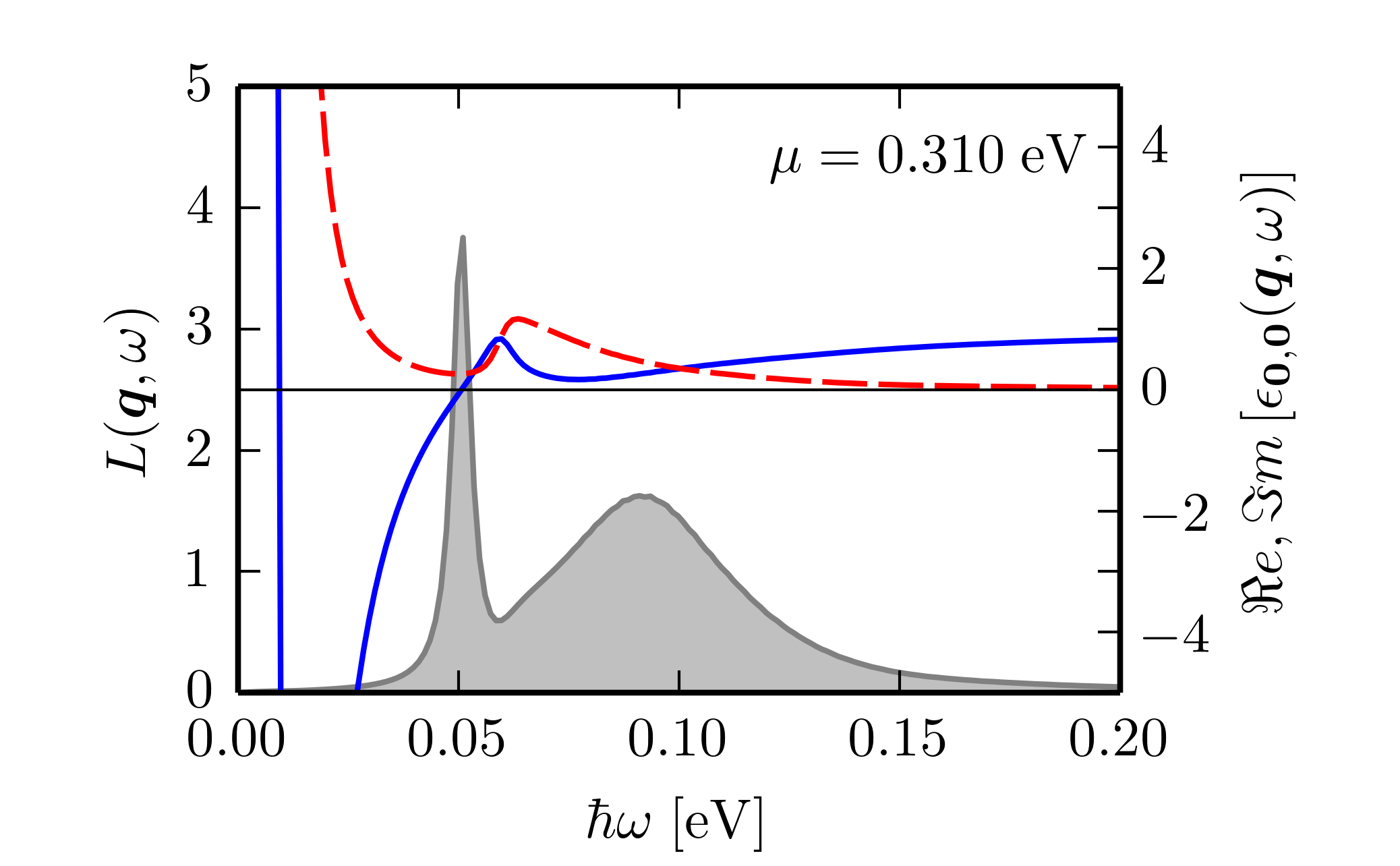}\put(2,60){b)}\end{overpic}
\begin{overpic}[width=3.46in]{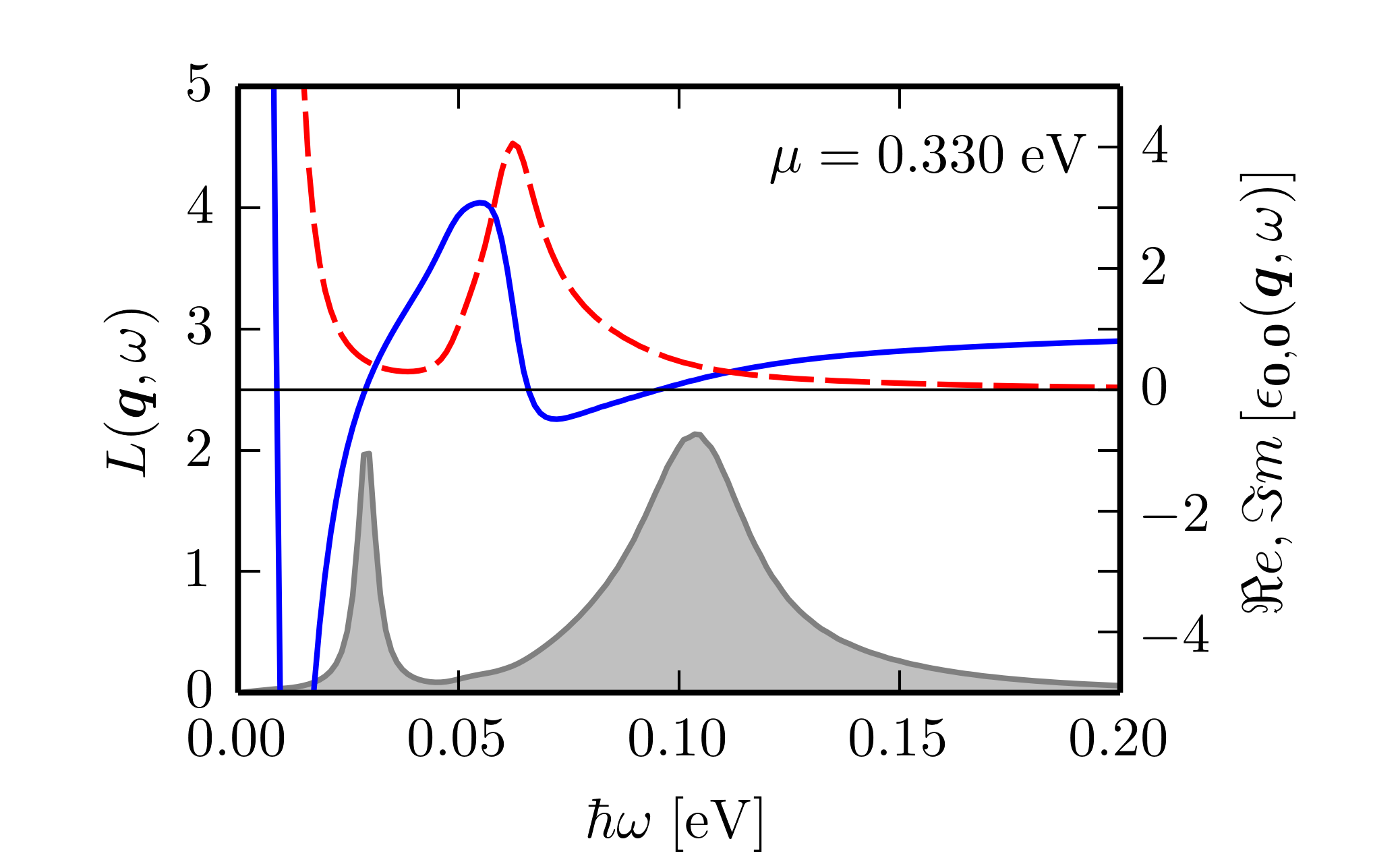}\put(2,60){c)}\end{overpic}
\begin{overpic}[width=3.46in]{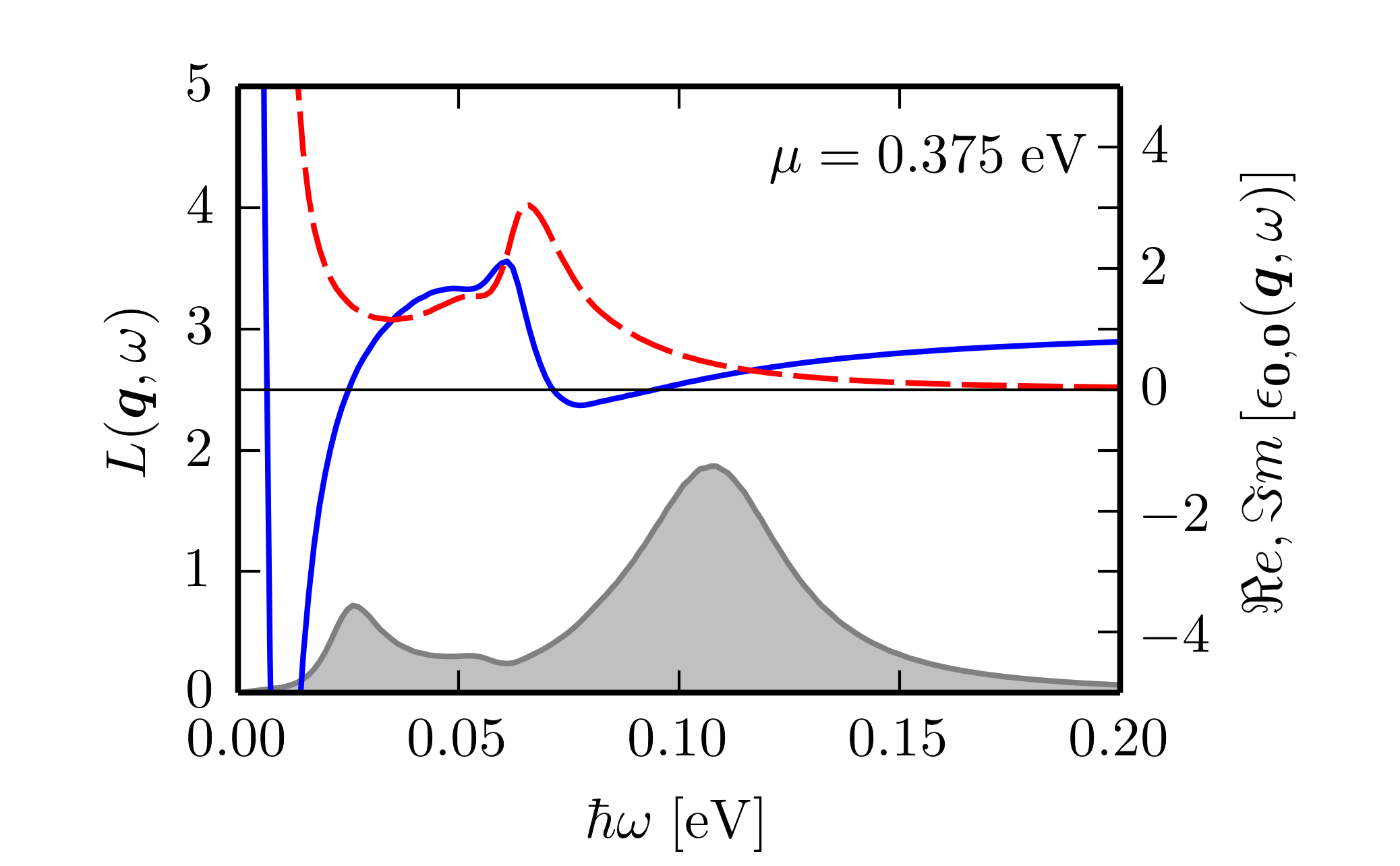}\put(2,60){d)}\end{overpic}
\caption{\label{fig:masspotential_appendix}
Energy dependence of the loss function (grey-shaded area), calculated for a moir\'e potential as 
in Fig.~\ref{fig:bands}c) of the main text.
The loss function is calculated at $|{\bm q}| = 0.021~{\rm nm}^{-1}$, as in Fig.~\ref{fig:masspotential} of the main text.
The panels from a) to d) correspond to $\mu = 0.250~{\rm eV}$, $0.310~{\rm eV}$, $0.330~{\rm eV}$, and $0.375~{\rm eV}$, respectively. Solid and dashed lines refer to the real and imaginary part of the macroscopic dielectric function $\epsilon_{{\bm 0}, {\bm 0}}({\bm q}, \omega)$, respectively.
The values of these functions can be read from the vertical axis on the right.}
\end{figure*}
\begin{figure}
\begin{overpic}[width=3.46in]{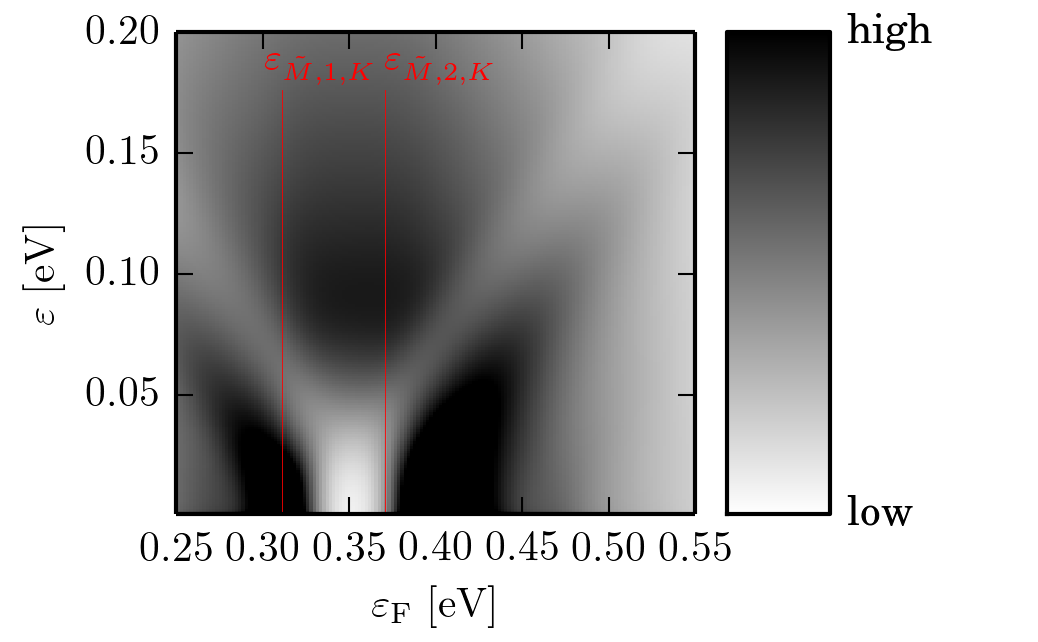}\put(2,60){a)}\end{overpic}
\begin{overpic}[width=3.46in]{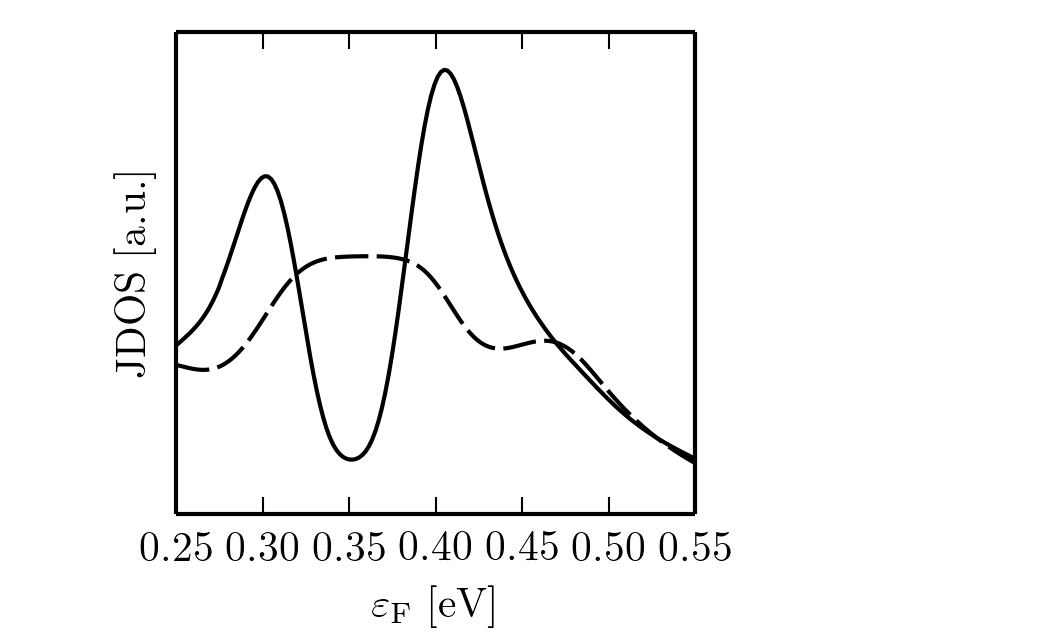}\put(2,60){b)}\end{overpic}
\caption{(Color online) JDOS (arbitrary units) as calculated from Eq.~(\ref{eq:jdos}) and from the density-of-states of the first and second conduction minibands in Fig.~\ref{fig:bands}c) of the main text. 
Panel a) shows the JDOS $J(\varepsilon)$ as a function of $\varepsilon$ and of the Fermi energy $\varepsilon_{\rm F}$. Panel b) shows the JDOS as a function of the Fermi energy $\varepsilon_{\rm F}$, 
for $\varepsilon = 25.0~{\rm meV}$ (solid line) and $\varepsilon = 100.0~{\rm meV}$ (dashed line). \label{fig:jdos}}
\end{figure}
\begin{figure}[t]
\begin{overpic}[width=3.46in]{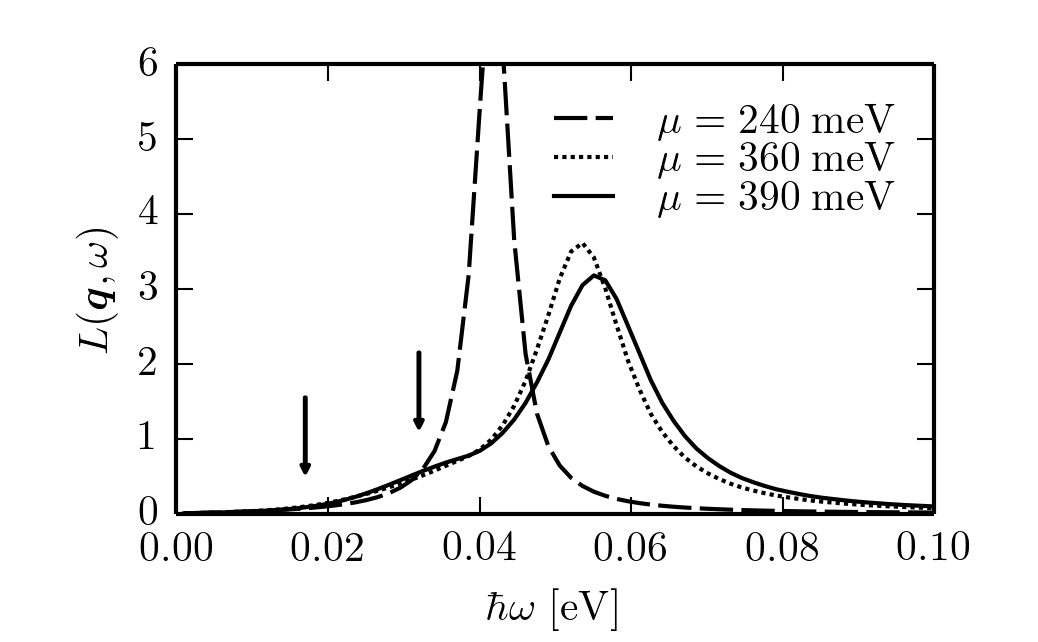}\put(2,60){a)}\end{overpic}
\begin{overpic}[width=3.46in]{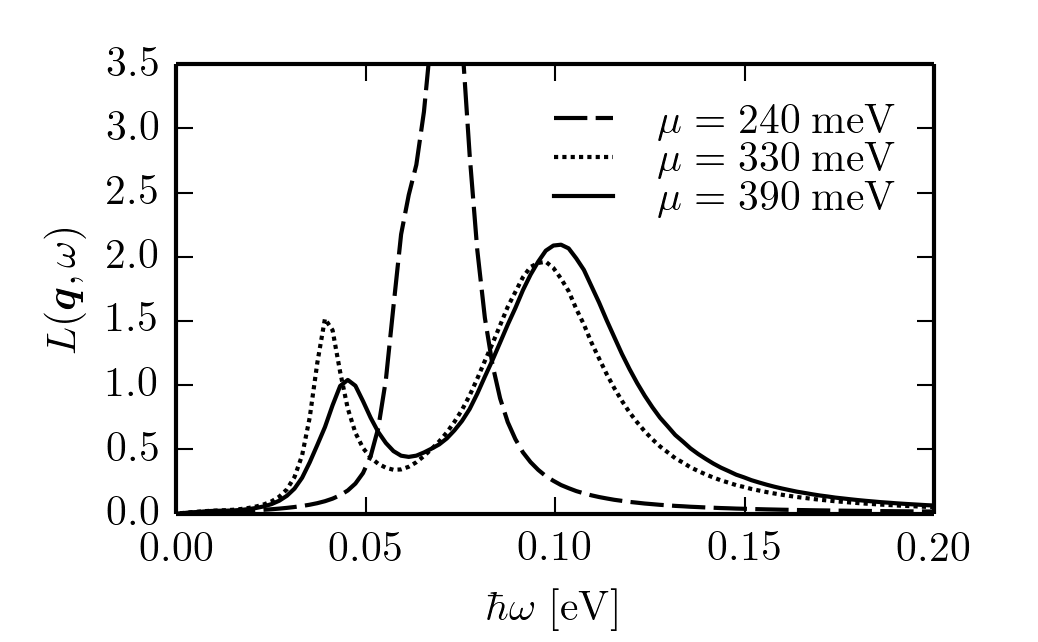}\put(2,60){b)}\end{overpic}
\caption{Energy and temperature dependence of the loss function $L({\bm q}, \omega)$.
In panels a) and b) we use the same parameters as in Figs.~\ref{fig:scalarpotential} and~\ref{fig:masspotential} of the main text, respectively. Data in this Figure have been calculated by using $T = 300~{\rm K}$, while Figs.~\ref{fig:scalarpotential} and~\ref{fig:masspotential} in the main text refer to $T = 10~{\rm K}$.
In panel a), the vertical arrows indicate the position of the satellite $\tilde{M}$-point plasmon for two different values of the chemical potential $\mu$ ($\mu =360~{\rm meV}$ and $\mu = 390~{\rm meV}$). This mode, which is clearly visible at $T = 10~{\rm K}$---see Fig.~\ref{fig:scalarpotential} in the main text---is severely broadened at $T = 300~{\rm K}$. \label{fig:temperature}}
\end{figure}

In Fig.~\ref{fig:scalarpotential_appendix}, we show the real and imaginary part of the macroscopic dielectric function $\epsilon_{{\bm 0}, {\bm 0}}({\bm q},\omega)$, together with the loss function $L({\bm q}, \omega)$, in the case of a scalar potential with $V_{\rm s} = 30~{\rm meV}$ and $V_{\Delta} = 0$. We remind the reader that this yields the miniband structure shown in Fig.~\ref{fig:bands}a) of the main text. Two values of the chemical potential $\mu$ are considered, below [panel a)] and above [panel b)] the upper edge of the first conduction miniband.
In panel a) only one zero of the real part of the dielectric function is present, corresponding to the standard DP~\cite{grapheneplasmonics} of a 2D MDF fluid.
In the second case, we notice three zeroes of the real part of the dielectric function. A self-sustained collective mode of an electron liquid corresponds~\cite{Giuliani_and_Vignale} to a zero $\omega_0$ of the real part of the dielectric function with $\partial \Re e [\epsilon_{{\bm 0}, {\bm 0}}({\bm q},\omega)]/\partial \omega|_{\omega = \omega_0} >0$. Zeroes such that $\partial \Re e [\epsilon_{{\bm 0}, {\bm 0}}({\bm q},\omega)]/\partial \omega|_{\omega = \omega_0} <0$ have to be discarded since they would correspond to collective modes whose amplitude grows in time~\cite{Giuliani_and_Vignale}, i.e.~they would be poles of the density-density linear response function located in the {\it upper} (rather than {\it lower}) half of the complex plane. In Fig.~\ref{fig:scalarpotential_appendix}b) we note the existence of a new zero with positive slope at energies $\lesssim 25~{\rm meV}$.
For the corresponding value of the chemical potential (i.e.~$\mu \simeq 0.37~{\rm eV}$), the bottom of the second conduction miniband---see Fig.~\ref{fig:bands}a) in the main text---hosts an electron pocket in the neighborhood of the $\tilde{M}$ point of the SBZ.
Consequently, we identify the novel zero of the real part of the dielectric function with a collective oscillation of the electrons in that pocket, and we denote it as a satellite ``$\tilde{M}$-point'' plasmon.
Similar notation is used throughout the main text for other modes arising from electrons or hole pockets in the neighborhood of any high-symmetry point in the SBZ. As discussed in the main text, a more accurate identification of satellite plasmons is obtained by calculating the effective carrier density of each pocket and fitting the plasmon dispersion. The effective carrier density in a pocket is further discussed in Appendix~\ref{app:dependence}.

In Fig.~\ref{fig:masspotential_appendix}, we show the same quantities as in Fig.~\ref{fig:scalarpotential_appendix} but for the case $V_{\rm s} = 0$ and $V_{\Delta} = 30~{\rm meV}$, corresponding to the minibands shown in Fig.~\ref{fig:bands}c) 
of the main text.
We see that the zero of the real part of the dielectric function, which corresponds to the ordinary DP for low values of the chemical potential $\mu$ [Fig.~\ref{fig:masspotential_appendix}a)], {\it continuously} shifts to lower energies when the chemical potential $\mu$ increases and approaches the upper edge of the first conduction miniband. The DP mode morphs into a satellite $\tilde{K}$-point plasmon, generated by a pocket of {\it holes} in the vicinity of the Dirac crossing between the first and second conduction minibands. In Figs.~\ref{fig:masspotential_appendix}b)-d) we also see a broad peak in the loss function at an energy $\hbar\omega \simeq 0.10~{\rm eV}$, which occurs close to a peak in the {\it imaginary} part of the dielectric function. Although in Figs.~\ref{fig:masspotential_appendix}b)-d) the real part of the dielectric function crosses zero around $\hbar\omega \simeq 0.10~{\rm eV}$ with the ``right'' (i.e.~positive) slope, we believe that the broad peak in the loss function should not be interpreted as a proper plasmon mode but, rather, as a peak arising from inter-band transitions between the first and second conduction minibands. Indeed, in the long-wavelength $q \to 0$ limit, 
this zero in the real part of the dielectric function disappears---see Fig.~\ref{fig:qdependence}b) in the main text. To further corroborate our interpretation, we have calculated the weighted joint density-of-states (JDOS) for optical transitions, for the first and second conduction minibands, which is defined as
\begin{equation}\label{eq:jdos}
J(\varepsilon) = \frac{1}{\varepsilon} \int_{\varepsilon_{\rm F} - \varepsilon}^{\varepsilon_{\rm F}} d\varepsilon' D(\varepsilon') D(\varepsilon' + \varepsilon)~,
\end{equation}
where $D(\varepsilon)$ is the standard density-of-states as a function of energy,
\begin{equation}
D(\varepsilon) = 2 \, \sum_{\nu} \sum_{n} \sum_{{\bm k} \in {\rm SBZ}} \delta(\varepsilon - \varepsilon_{{\bm k}, n, \nu})~.
\end{equation}
In Fig.~\ref{fig:jdos} we see that for energies $\varepsilon \lesssim 50~{\rm meV}$ the JDOS has a minimum, roughly corresponding to the gap at the $\tilde{M}$ point of the SBZ. This minimum is accompanied 
by two maxima at Fermi energies $\varepsilon_{\rm F} \simeq 30~{\rm meV}$ and $\varepsilon_{\rm F} \simeq 
40~{\rm meV}$. These maxima in the JDOS stem from a saddle point and a minimum in the first and second conduction minibands, respectively.
At higher energies ($\varepsilon \simeq 100~{\rm meV}$) the JDOS displays a broad peak, which stems from inter-band transitions. This therefore confirms our interpretation of the broad peak in the loss function at $\hbar\omega \simeq 100~{\rm meV}$ as originating from inter-band particle-hole excitations.

Before concluding this Appendix we discuss the role of temperature. In Fig.~\ref{fig:temperature}, we illustrate the dependence of the satellite plasmon modes on temperature.
The results presented in the main text refer to $T = 10~{\rm K}$. Here we use room temperature, $T = 300~{\rm K}$.
We notice that the DP mode, i.e.~the main peak in Fig.~\ref{fig:temperature}, is slightly affected by the temperature increase; the satellite $\tilde{M}$-point plasmon in panel a) is instead quite sensitive to temperature and very broad at $T=300~{\rm K}$. Its spectral weight is anyway responsible for a visible shoulder in the loss function for energies lower than those of the DP. In panel b), instead, we clearly see that both the $\tilde{K}$- and $\tilde{M}$-point plasmons (visible at $\mu = 330~{\rm meV}$ and $\mu = 390~{\rm meV}$, respectively) are still quite sharp at room temperature.

\section{The dependence of $n_{\kappa, \nu}$ on $\mu$}
\label{app:dependence}
\begin{figure*}
\begin{overpic}[width=3.46in]{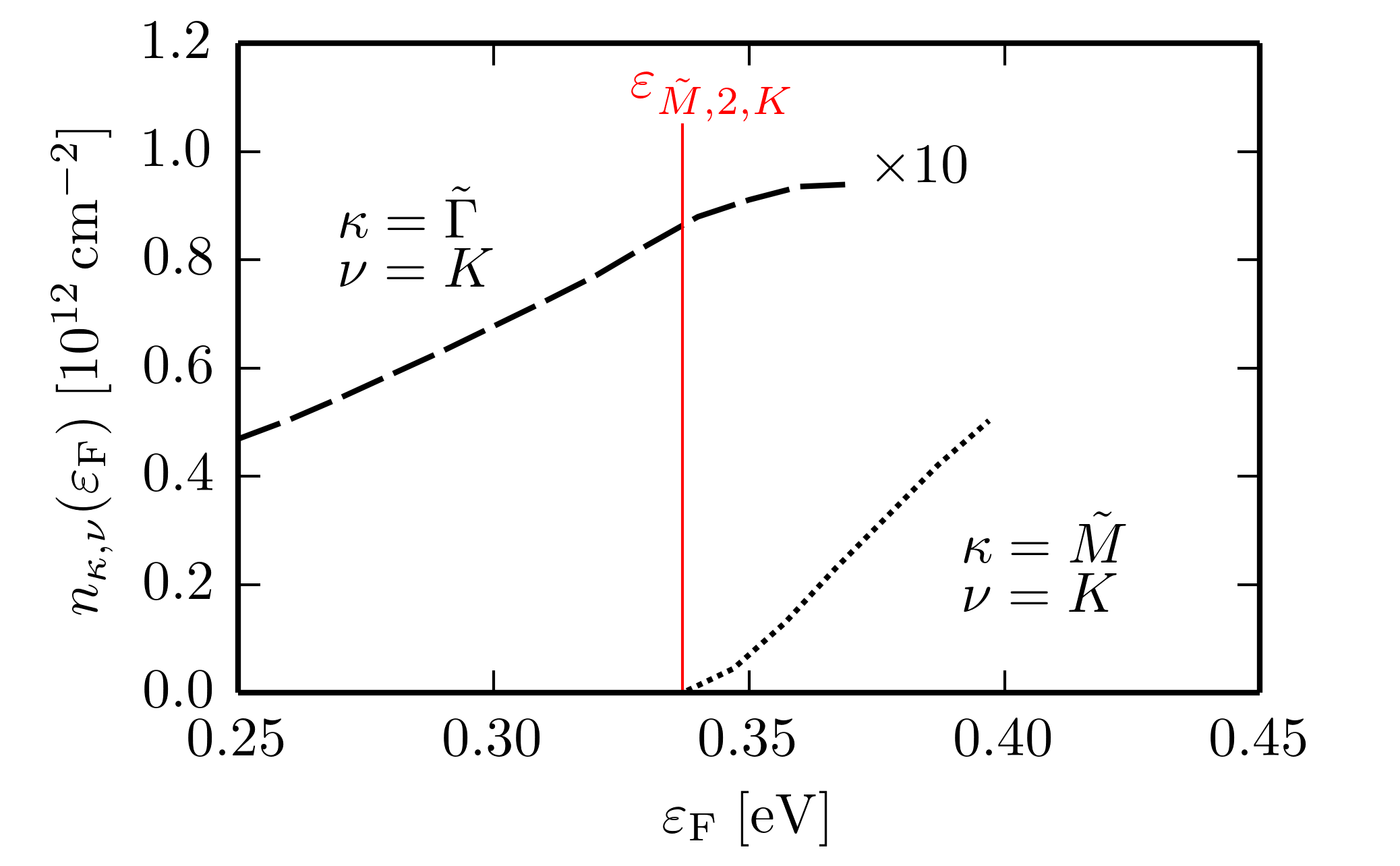}\put(2,60){a)}\end{overpic}
\begin{overpic}[width=3.46in]{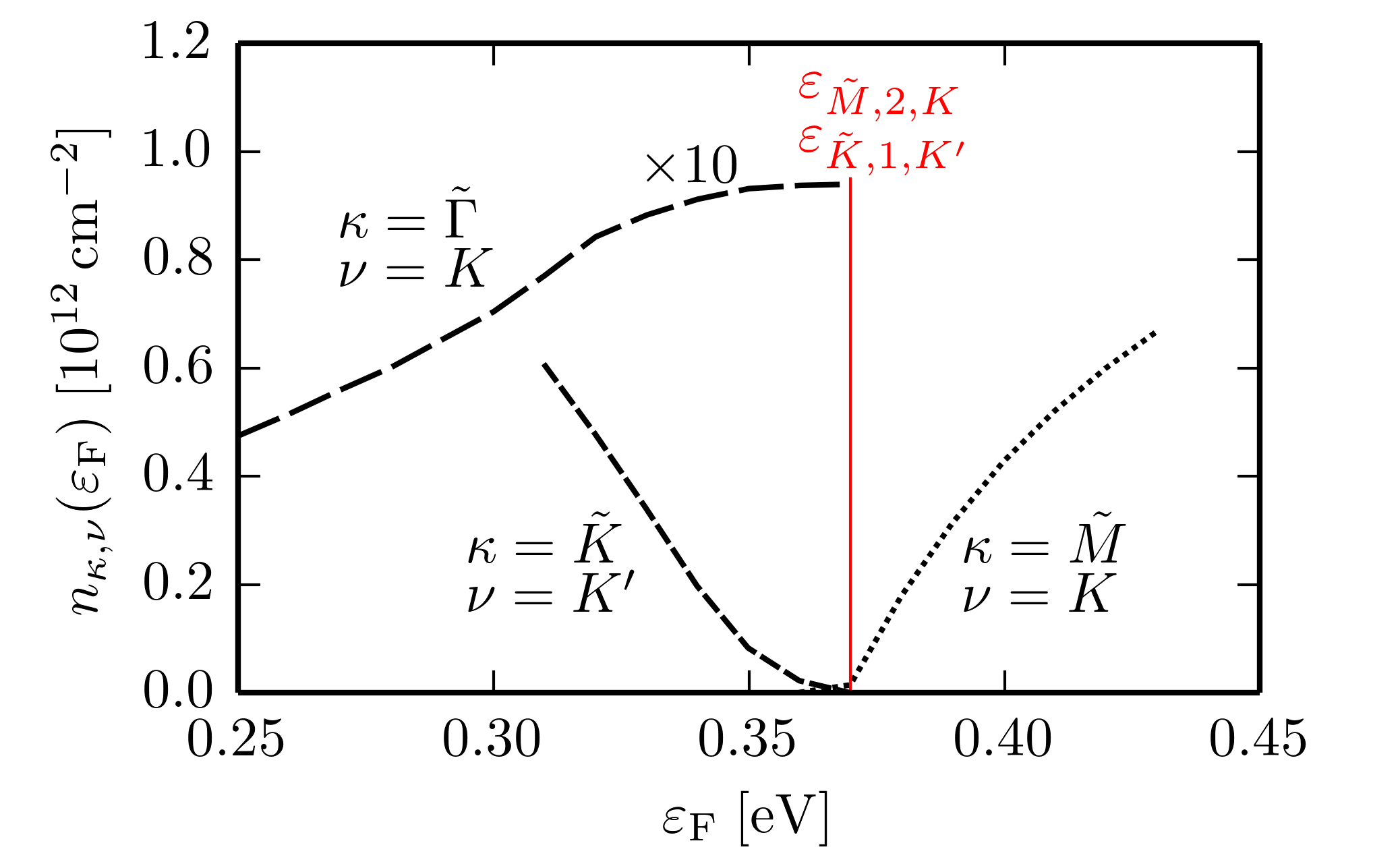}\put(2,60){b)}\end{overpic}
\caption{(Color online) Electron and hole densities, hosted in pockets located around the high-symmetry points in the SBZ, as functions of the Fermi energy $\varepsilon_{\rm F}$ at $T=0$. Data in panel a) [panel b)] have been calculated for the same parameters as in Fig.~\ref{fig:bands}a) [Fig.~\ref{fig:bands}c)] in the main text.
In both panels, the dashed line corresponds to the electron density in the first conduction band in the $K$ valley, which has been divided by a factor $10$ to fit into the frame of the figure. The dotted line corresponds to the electron density in the neighborhood of the $\tilde{M}$ point in the SBZ, in the $K$ valley.
In panel b), the density of the electron pocket vanishes linearly at the band edge (vertical solid line), as expected for a parabolic band minimum, while the behaviour is less clear in panel a), where a saddle point is present in the band dispersion.
In panel b), the short-dashed line corresponds to the hole density in the neighborhood of the $\tilde{K}$ point in the SBZ, in the $K'$ valley.
The density of the hole pocket vanishes quadratically at the band edge (vertical solid line) as expected for a Dirac-type crossing. These densities are used in the calculation of the dependence of the plasmon frequency on chemical potential, which is shown in Figs.~\ref{fig:scalarpotential} and~\ref{fig:masspotential} in the main text. \label{fig:density}}
\end{figure*}

In the case of the minibands shown in Fig.~\ref{fig:bands}a) in the main text, it is easy to identify the origin of the satellite $\tilde{M}$-point plasmon because only one electron pocket appears at the crossing between the first and second conduction minibands.
However, in the more complicated case shown in Fig.~\ref{fig:bands}c), two electron pockets appear, at the $\tilde{M}$ and $\tilde{K}$ points in different principal valleys (and in the corresponding equivalent points in the SBZ). 
A more quantitative analysis of the dependence of the plasmon frequency on chemical potential is therefore necessary.
This consideration is based on the fact that a pocket with parabolic dispersion---as in the case of the bottom of the second conduction miniband at the $\tilde{M}$ point in one principal valley, Fig.~\ref{fig:bands}c)---is expected to generate a standard 2DEG-type plasmon~\cite{Giuliani_and_Vignale}, while a pocket located around a linear band crossing---as in the case of the Dirac crossing between the first and second conduction minibands at the $\tilde{K}$ point in the other principal valley---is expected to generate a DP~\cite{grapheneplasmonics,Diracplasmons}.
The frequency of these two kinds of plasmons scales differently with the carrier density $n$ in the pocket, i.e.~$\propto \sqrt{n}$ in the 2DEG case and $\propto n^{1/4}$ in the DP case. We therefore need to calculate the carrier density in each pocket as a function of the chemical potential $\mu$, which is shown in Fig.~\ref{fig:density} at $T = 0$, and use this information in the analytical long-wavelength formulas for the plasmon dispersions reported in the main text. Each formula has a single fitting parameter: the effective mass in the case of a parabolic-type dispersion and the Fermi velocity in the case of a linear dispersion.
We reiterate that, in the explored range of parameter space, no evidence for anisotropies in the plasmon dispersion has emerged---this explains why a single parameter is sufficient to well fit the numerical data extracted from the sharp peaks in the loss function.


\begin{thebibliography}{77}
\bibitem{footnote_asymmetry}
	Following Ref.~\onlinecite{wallbank_prb_2013},
	we assume that inversion-asymmetric terms (which are allowed for graphene on hBN)
	have negligible magnitude.
\end{thebibliography}
\end{document}